\documentclass[aps,prb,twocolumn,superscriptaddress,showpacs,footnotebib]{revtex4-2}
\usepackage{amsmath}
\usepackage{amssymb}
\usepackage{physics}
\usepackage{bm}
\usepackage{color}
\usepackage{graphicx}
\usepackage{epstopdf}
\usepackage{epsfig}
\usepackage{amsfonts}
\usepackage[naturalnames]{hyperref}
\usepackage{hypcap}
\usepackage{verbatim}
\usepackage{tabularx}
\usepackage{bbm}
\usepackage{esvect}
\usepackage{subfigure}
\usepackage{slashed}
\usepackage[export]{adjustbox}
\usepackage{tikz}
\usetikzlibrary{quantikz}
\usepackage{amsthm}
\newtheorem*{theorem}{Theorem}

\def\bea{\begin{eqnarray}}
\def\eea{\end{eqnarray}}
\def\nn{\nonumber}
\def\ba{\begin{array}}
\def\ea{\end{array}}
\def\Tr{\text{Tr}}
\def\nn{\nonumber}

\def\Tr{\text{Tr}}
\newcommand{\kett}[1]{|#1\rangle\rangle}
\newcommand{\braa}[1]{\langle\langle#1|}
\newcommand{\kettid}{|\text{id}\rangle\rangle}
\newcommand{\braaid}{\langle\langle \text{id}|}
\newcommand{\kettswap}{|\text{swap}\rangle\rangle}
\newcommand{\braaswap}{\langle\langle \text{swap}|}

\hypersetup{colorlinks=true, citecolor=blue, urlcolor=blue, linkcolor=blue}
\begin{document}
\title{Phase transitions in sampling and error correction in local Brownian circuits}

\author{Subhayan Sahu}
\email{ssahu@perimeterinstitute.ca}
\affiliation{Perimeter Institute for Theoretical Physics, Waterloo, Ontario N2L 2Y5, Canada}
\author{Shao-Kai Jian}
\email{sjian@tulane.edu}
\affiliation{Department of Physics and Engineering Physics, Tulane University, New Orleans, Louisiana, 70118, USA}

\date{\today}

\begin{abstract}
We study the emergence of anticoncentration and approximate unitary design behavior in local Brownian circuits. 
The dynamics of circuit averaged moments of the probability distribution and entropies of the output state can be represented as imaginary time evolution with an effective local Hamiltonian in the replica space. 
This facilitates large scale numerical simulation of the dynamics in $1+1d$ of such circuit-averaged quantities using tensor network tools, as well as identifying the various regimes of the Brownian circuit as distinct thermodynamic phases. 
In particular, we identify the emergence of anticoncentration as a sharp transition in the collision probability at $\log N$ timescale, where $N$ is the number of qubits. 
We also show that a specific classical approximation algorithm has a computational hardness transition at the same timescale. 
In the presence of noise, we show there is a noise-induced first order phase transition in the linear cross entropy benchmark when the noise rate is scaled down as $1/N$. 
At longer times, the Brownian circuits approximate a unitary 2-design in $O(N)$ time. 
We directly probe the feasibility of quantum error correction by such circuits, and identify a first order transition at $O(N)$ timescales. 
The scaling behaviors for all these phase transitions are obtained from the large scale numerics, and corroborated by analyzing the spectrum of the effective replica Hamiltonian.
\end{abstract}

\maketitle

\tableofcontents

\section{Introduction}

Random quantum circuits (RQC) play a pivotal role in both quantum dynamics theory and quantum information theory, offering insights into fundamental aspects such as quantum chaos, out-of-time correlation functions, and entanglement entropy~\cite{nahum2018operator,von2018operator,zhou2019emergent,khemani2018operator,rakovszky2018diffusive}. 
Closely related to RQC, random tensor network serves as a valuable tool for investigating the AdS/CFT correspondence, a theory aiming to understand quantum gravity through quantum entanglement~\cite{hayden2016holographic}. 
Additionally, random quantum circuits find extensive applications in quantum information theory, including quantum advantage~\cite{arute2019quantum, morvan2023phase, wu2021strong, zhu2022quantum}, quantum error correction~\cite{Brown_2013,Brown_2015, Gullans_QuantumCoding_2021}, etc.

Random circuits are expected to be a toy model capturing the following properties of generic quantum circuits: they output states of high complexity and generate maximal entanglement between initially disconnected regions. 
An important question is characterizing the time (depth) required for achieving the high complexity. 
How do we characterize the complexity of random circuits? In this work, we focus on two distinct features: anticoncentration and unitary design. 

\begin{figure}[htp]
    \centering
    \includegraphics[width = \columnwidth]{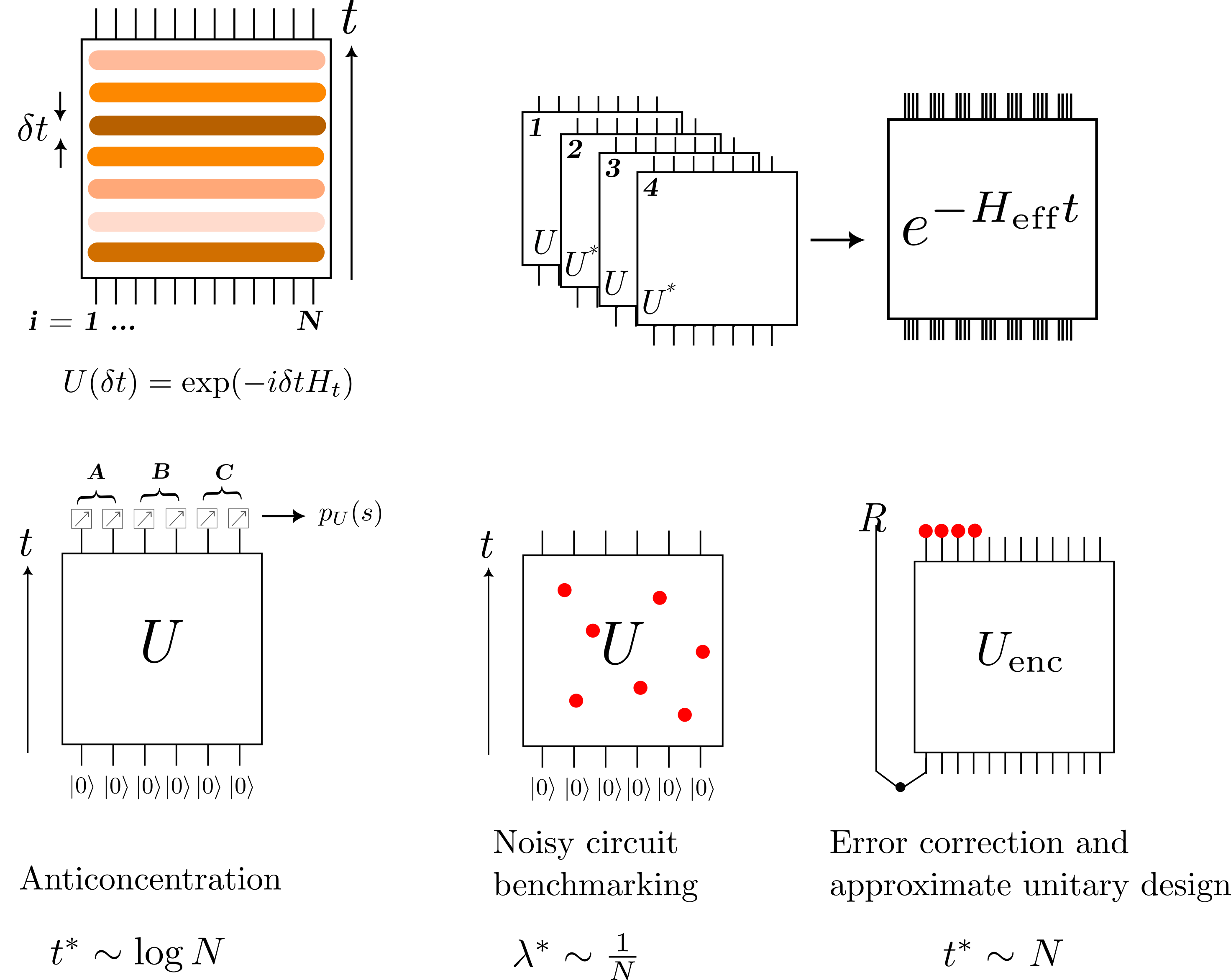}
    \caption{We consider local Brownian circuits acting on qubits on a chain. 
    Averaging the replica dynamics $U\otimes U^{*}\otimes U \otimes U^{*}$ leads to an imaginary time evolution with a local Hamiltonian in replica space, $H_{\text{eff}}$. 
    We can simulate this imaginary time evolution in $1+1d$ efficiently with tensor networks. 
    We study three scenarios: anticoncentration, noisy circuit benchmarking, and approximate unitary design (and consequently, error correction), and find distinct quantum informational phases at different time and noise regimes, separated by phase transitions. 
    These results are corroborated using large scale numerics and analyzing the spectrum of $H_{\text{eff}}$.}
    \label{fig:summary}
\end{figure}

Consider a circuit $C$ acts on an initial simple state (the product state of $\ket{0}$ on all qubits) and the output state is measured in the computational basis to obtain a distribution over measurement outcomes, $p_{C}(s) = \left|\bra{s}C\ket{0}\right|^{2}$. 
Anticoncentration is the property that $p_{C}(s)$ is well spread over all bitstrings $s$. 
Certifying that the circuit is anticoncentrated is crucial in guaranteeing that the RQC simulation is classically hard, and is a promising route towards demonstrating quantum advantage~\cite{Boixo_2018, hangleiter2023computational, arute2019quantum,wu2021strong, zhu2022quantum}. 

At long enough depths, RQC has a stronger notion of complexity: it becomes an approximate unitary design. 
A unitary ensemble is said to be $k-$design if it approximates a global Haar random unitary in its first $k$ moments. 
In particular, ensuring that a RQC has achieved the 2-design property is enough for the RQC to be maximally decoupling. 
Consider a system $A$, initially maximally entangled with a reference $R$, is subjected to a circuit $C$, before being coupled to an environment $E$. 
The initial encoding via $C$ is said to have the decoupling property if the joint density matrix on $R\cup E$ is approximately factorizable $\rho_{RE} \approx \rho_{R}\otimes \rho_{E}$. 
This can also be associated with the RQC dynamically generating a quantum error correcting code~\cite{Brown_2013,Brown_2015, Gullans_QuantumCoding_2021}.

Several avenues of research on RQC have established that anticoncentration and unitary design occur at parametrically distinct timescales. 
Suppose we consider circuits with spatial local connectivity in $d$ dimensions. 
Past research has shown that ensembles of RQC with Haar random local gates achieve anticoncentration and unitary design in $O(\log N)$~\cite{Dalzell_2022} and $O(N^{1/d})$~\cite{Brandao_2016_local,Harrow2023approximate,hunterjones2019unitary} timescales, respectively, where $N$ denotes the number of qubit.
Note that both anti-concentration and 2-design property are diagnosed by non-linear properties of the quantum state generated by the circuits. 
This makes numerically simulating these properties for local Haar random circuits hard and limited to modest system sizes and for short times. 
Hence, much of the research on RQC has depended on proving analytical bounds, classically simulable Clifford circuits, and perturbations around semi-classical limits, such as large local Hilbert space dimensions.

In this work, we provide a minimal model which allows us to do efficient and guaranteed numerical simulation of the quantum informational quantities probing anticoncentration and 2-design property of large-sized random circuits using tensor network technology. 
We take the approach of directly representing the informational quantities averaged over the circuit ensemble as a linear observable in a replicated Hilbert space. 
Here, replicas are simply exact copies of the original system, and the informational observables probe the correlation between different replicas. 
We study a particular ensemble of RQC, namely local Brownian circuits~\cite{Lashkari_2013, Onorati_2017, Bentsen_2021, Sahu_2022, jian2022linear}.

These Brownian qubit models can be defined in any graph where each vertex hosts a qubit, with nearest neighbor Brownian interaction generating the unitary evolution. 
Remarkably, the real-time evolution of circuit averaged non-linear observables of the density matrix can now be realized as imaginary time evolution in the replica space. 
The Hilbert space for $k$ replicas is simply the combination of a forward contour and a backward contour for real-time evolution for each replica; so the local Hilbert space encompasses $2k$ spins.  
After averaging of Brownian couplings, the quantum dynamics reduces to a Hermitian replica qubit Hamiltonian with the same locality properties as the initial interaction graph. 
This model not only establishes a clear mapping between various quantum information quantities and those of a quantum spin model, but also transforms the problem of quantum dynamics into a thermodynamic problem.

Furthermore, imaginary time evolution with local Hamiltonians is guaranteed to be efficient in 1$d$ using simple matrix product state and Time Evolving Block Decimation (TEBD) algorithms~\cite{Vidal_2003,White_2004,Daley_2004}. 
This allows us to perform large-scale simulations of these informational quantities in $1+1$ dimensional circuit. 
As an example, we can simulate the averaged R\'enyi-2 entanglement properties of a Brownian circuit on $N\sim O(100)$ qubits for $t\sim O(N)$ depths in a few minutes on a standard laptop. 

The effective Hamiltonian approach also provides a statistical mechanical description of different regimes of RQC as distinct `phases', separated by phase transitions. 
These phases can be described within a generalized Landau framework involving multiple replicas, where the relevant symmetry is the replica permutation symmetry~\cite{jian2021phase} (when we introduce multiple identical copies of the system, they can be permuted amongst each other without changing the effective description). 
Specifically, in the two replica scenario that we focus on in this work, the effective Hamiltonian has a $\mathbb{Z}_{2}$ symmetry corresponding to a relative swap between the two real-time contours, which turns out to be the relevant symmetry for non-linear observables of the density matrix. 
This effective Hamiltonian is essentially a $\mathbb{Z}_{2}$ Ising model in the replica space, and the phases of quantum information and their phase transitions are associated with the various phases and critical properties of this Ising model.

%In fact, the informational phases can be identified from the interplay of the symmetry of the effective Hamiltonian, which is essentially a , and the boundary conditions determining the exact observable we are interested in. This naturally implies that the distinct regimes of anti-concentration and unitary designs can be associated with different order-disorder phenomena in the effective Hamiltonian picture.

Using large scale numerics of the Brownian circuit model we can directly probe the dynamical properties of the quantum informational quantities, and identify the saturation to anticoncentration (at $\sim \log N$ depth) and 2-design property (at $\sim N$ depth) of the RQC as sharp transitions, confirmed by careful finite-size scaling of the numerical data. 
This can be understood analytically by investigating the spectral properties of the effective Hamiltonian. 
The anticoncentration transition can also be directly associated with a transition in the computational hardness of classically simulating the output distribution. 
To show this, we show that a specific algorithm for simulating the output distribution~\cite{Napp_2022} undergoes a hardness transition in $\sim \log N$ depth.

In order to study the 2-design transition, we focus on investigating the feasibility of the Brownian circuit as a quantum error-correcting code, by directly simulating a quantity akin to the mutual information between the reference and environment in the decoupling setup, named `Mutual Purity'~\cite{balasubramanian2023quantum}. 
The mutual purity is a 2-replica quantity, and has recently been shown to provide a bound for the error correction capabilities of RQC in~\cite{balasubramanian2023quantum}. 
We show that the mutual purity undergoes a first order transition in $O(N)$ time, after which the Brownian circuit approximates the global Haar random unitary for coding purposes. 
This coding transition is a first order pinning transition, driven by boundary conditions determining the mutual purity, akin to~\cite{Gullans_QuantumCoding_2021}. 
Furthermore, the mutual purity contributes to a bound for the failure probability for correcting errors after the encoding by the RQC. 
By numerically computing the mutual purity for different error models after the 2-design transition, we can also find a first order threshold transition for the code distance.

As mentioned earlier, sampling of RQC outcome states is one of the most promising routes towards demonstration of quantum advantage in near term quantum devices~\cite{Boixo_2018,arute2019quantum,wu2021strong,zhu2022quantum,hangleiter2023computational}. 
However, real quantum devices suffer from noise. 
In order to benchmark that the noisy quantum device, an estimate of the fidelity of the output state is desirable. 
One proposal for an efficient estimate for the fidelity is the linear cross-entropy benchmark $\chi_{\text{XEB}}$, and a high score in this benchmark suggests that the RQC simulation is classically hard~\cite{Boixo_2018}. 
However, it has recently been understood that with local noise models, there is a noise-induced phase transition (NIPT) in the linear cross entropy benchmarking~\cite{morvan2023phase,ware2023sharp,dalzell2021random, Aharonov_2023}. 
In the weak noise regime, $\chi_{\text{XEB}}$ provides a reliable estimate of fidelity, and in the strong noise regime, it fails to accurately reflect fidelity. 
Furthermore, this implies that in the strong noise regime, classical simulation can yield a high score in the cross-entropy benchmark~\cite{barak2020spoofing,gao2021limitations}, without necessarily solving the sampling task. 
The noise model can be incorporated in our Brownian circuit setup, where the noise serves as an explicit replica-permutation symmetry breaking field~\cite{jian2021quantum}. 
Using a combination of numerical and analytical tools, we characterize the NIPT in benchmarking by identifying it as a first order phase transition in the effective Hamiltonian picture.  

%When RQC are combined with measurements, a remarkable phenomenon occurs when the measurement frequency is tuned: we get a continuous measurement -induced phase transition (MIPT) in the entanglement structure of the steady state conditioned on measurement outcomes~\cite{li2018quantum,li2019measurement,skinner2019measurement,chan2019unitary,gullans2020dynamical}. 
%In the most well-studied scenario, the entanglement entropy of the steady state undergoes a transition from volume law to area law. 
%Using the Brownian circuit model, we can easily examine the corresponding MIPT in the R\'enyi-2 entropy. 
%Specifically, we investigate the phases of our effective Hamiltonian by incorporating weak measurements with post-selection. 
%Unlike noise, which is modeled by a symmetry-breaking field, measurements do not break the replica permutation symmetry. 
%Rather, measurements are akin to a transverse field within the effective quantum spin model in the replica space. Consequently, the transition observed in our quantum Ising model precisely corresponds to a $2-d$ Ising transition with a central charge of $c=1/2$ for the 2 replica case, which we confirm by density matrix renormalization group (DMRG) calculation with the effective Hamiltonian. 

%\R{Comparison of local Brownian circuit with local Haar Random circuits}

\subsection{Main results and outline of paper}
\begin{table}[htp]
\includegraphics[width = \columnwidth]{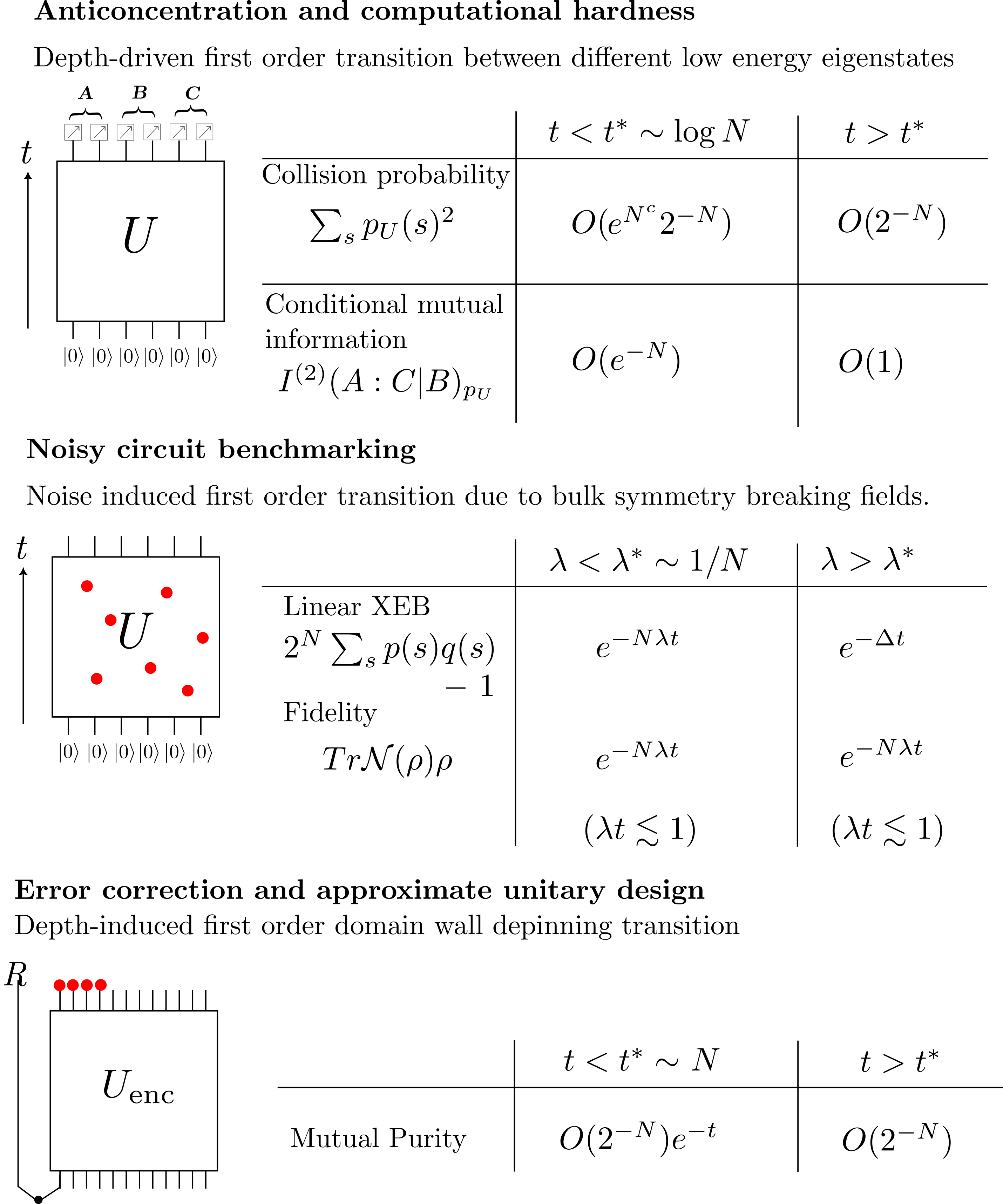}
\caption{The transition point, nature of the transition, and the asymptotic scaling for various information quantities for the Brownian circuit are shown in this table.\label{table:summary}}
\end{table}

We first briefly summarize the results of the paper. 
The main results of the paper are represented in Fig.~\ref{fig:summary} and Table~\ref{table:summary}.

\begin{itemize}
    \item \textbf{Anticoncentration:} We probe anticoncentration in the $1+1$d Brownian circuit $U$ by computing the `collision probability'~\cite{Dalzell_2022}, defined as the circuit averaged probability that two independent samples of the RQC (acting on the $\ket{0^{\otimes N}}$ state of $N$ qubits) produce the same result, defined as $Z = \mathbb{E} \sum_{x}\left|\bra{x}U\ket{0^{\otimes N}}\right|^{4}$, where the averaging $\sim \mathbb{E}$ is done over all realizations of the circuit. 

    In the context of the effective Ising Hamiltonian ($H_{\text{eff}}$) description, we demonstrate that $Z$ equates to the transition probability between an imaginary-time evolved state from the initial state and a quantum paramagnetic state (defined in a later section). 
    The imaginary time evolution gradually projects the initial state onto the ground state of $H_{\text{eff}}$, which corresponds to $Z\sim 2^{-N}$. 
    However, in finite time $t$, excited states contributions result in $Z = 2^{-N} + S_\Delta e^{- \Delta t}$, where $\Delta$ ($S_\Delta$) denotes the energy gap (entropy) of the excitation~\footnote{In a one-dimensional chain with local couplings, the elementary excitation manifests as a domain wall, with a finite gap independent of the system size and an entropy proportional to the system size}. 
    The anticoncentration transition, thus occurs at $t = \frac{1}{\Delta} \log S_\Delta \sim \log N$, representing a depth-induced computational transition. 
    The $\log N$ results arise from the nature of the elementary excited states of the Ising model, and can be confirmed by direct large scale simulation of the imaginary time evolution.
    
    \item \textbf{Computational Hardness transition:} We probe the computational hardness of classically simulating the probability distribution in the measurement outcome in the earlier setup, i.e. $p_{U}(x) = \left|\bra{x}U\ket{0^{\otimes N}}\right|^{2}$. 
    By studying the R\'enyi-2 version of conditional mutual information (CMI) of $p_{U}(x)$ using numerics of the imaginary time evolution, we probe the hardness of a specific classical algorithm (`Patching algorithm') for approximately simulating the output distribution as introduced in~\cite{Napp_2022}. 
    We find that the CMI undergoes a phase transition at $O(\log N)$ time, with the same scaling behavior as the collision probability, signalling a computational hardness phase transition at the same depth.

    \item \textbf{Phase transition in cross-entropy benchmarking of noisy Brownian circuits:} Here we consider the following setup of two copies of the Brownian circuit, one that is affected by noise (denoted by the noisy channel $\mathcal{N}$), and the other copy undergoes the noise-free Brownian circuit. 
    We can now update the effective Hamiltonian with explicit noise in one of the replicas, $H_{\text{eff}} \to H_{\text{eff}}^{\prime}$. We can compute the fidelity $F = \Tr [\mathcal{N}(\rho)\rho] $ of the noisy simulation by doing imaginary time evolution with $H_{\text{eff}}^{\prime}$ (with local noise models, $H_{\text{eff}}^{\prime}$ remains local). We also compute the linear cross entropy benchmark, defined as $\chi_\text{XEB} = 2^N \sum_x p(x) q(x) - 1$, where $p(x)$ and $q(x)$ represent the output distribution in the noise-free and the noisy cases respectively~\cite{Boixo_2018}.

    In $H_{\text{eff}}^{\prime}$, noise explicitly breaks the Ising symmetry and subsequently pins the Ising spins. 
    Consider a local (unital) noise model, with $\lambda$ strength for each qubit (to be explicitly defined later). 
    Noise generically undermines the ferromagnetic phase that leads to anticoncentration, and leads to erosion of the quantum advantage. 
    This holds true for constant rate noise $\lambda \sim O(1)$. 
    Through the mapping to the quantum Ising model, we discover that noise behaves as a relevant perturbation with a scaling dimension of one. 
    Therefore, when the noise rate scales inversely with respect to the size of the chain, $\lambda \sim 1/N$, we get a noise-induced computational transition at some critical $\lambda^{*}\sim O(1/N)$. 
    This transition essentially resembles a field-induced first-order transition and conforms to finite size scaling with $\nu = 1/2$, which we confirm numerically. 
    Moreover, if the rate scales less (greater) than $1/N$, the noise is deemed irrelevant (relevant). 
    This result is consistent with recent results on Noise-induced phase transitions in cross entropy benchmarking~\cite{morvan2023phase,dalzell2021random,ware2023sharp}.

    We also study whether this transition signals a transition in the computational hardness in the simulation of noisy Brownian circuits. 
    By studying the R\'enyi-2 CMI of $p_{\mathcal{N}(U)}(x)$, we find that it does not undergo a hardness transition with depth for large enough depths, and actually exponentially decays with time. This suggests that the $1+1$d noisy random circuits are efficiently simulable in the long-time limit, even in the presence of infinitesimal scaled noise.

    \item \textbf{Coding transitions:} We encode some local information (a reference qubit $R$) in the entire system $A$ using the Brownian circuit, and probe the effectiveness of this encoding as a quantum error correcting code. 
    After encoding, the state on $A$ is affected by noise, which can be identified as a unitary coupling with the environment $E$. 
    Mutual purity $\mathcal{F}_{RE}$~\cite{balasubramanian2023quantum} is a two replica quantity which upper bounds the trace distance between the initial encoded state and the error affected encoded state after error correction using a recovery channel~\cite{Schumacher_1996,schumacher2001approximate}. 

    From the effective Hamiltonian perspective, the mutual purity can be represented as a transition probability between two ferromagnetic states. 
    In particular, we find that at short times the mutual purity decays exponentially, which can be identified with domain wall configurations pinned between the initial and final states; while after $t \sim O(N)$ time the domain walls get depinned, resulting in the saturation of the Mutual Purity to a global Haar value, i.e. realizes an approximate 2-design. 
    Using large-scale numerics, we are able to directly probe this transition to a 2-design as a first order depinning transition. 
    Furthermore, since the mutual purity determines the feasibility of error correction after the application of noise, we find a first order threshold transition in the fraction of qubits which are affected by noise. 
    The critical fraction can be identified as a lower bound for the `code distance' of the Brownian circuit as a quantum error correcting code.  

    %\item \textbf{Measurement-induced phase transition:} Interspersing the Brownian circuit with weak measurements leads to a modification of the effective Hamiltonian $H_{\text{eff}}$, where the measurements act as a transverse field in the quantum Ising model. In particular, if we focus on a particular quantum trajectory with postselection, the steady state of such a circuit can be identified with the ground state of $H_{\text{eff}}$. $H_{\text{eff}}$ undergoes a quantum phase transition, as the transverse field, i.e. the frequency of measurements, is tuned. We characterise this measurement-induced entanglement transition in the steady state using DMRG on the effective Hamiltonian, and find that the criticality of the R\'enyi-2 entropy agrees with Ising criticality in $2d$.

\end{itemize}

The paper is organized as follows. 
Section~\ref{sec:model} presents an introduction to the Brownian circuit model and a derivation of the effective Hamiltonian for $k=2$ replicas. 
In section~\ref{sec:symmetries}, we describe the symmetries of the effective Hamiltonian, and provide heuristic description of the phase diagrams. 
In section~\ref{sec:anticonc} we discuss the anticoncentration and computational hardness transition. 
In section~\ref{sec:noise} we investigate the noise-induced phase transition in benchmarking noisy Brownian circuits. 
In section~\ref{sec:qecc} we study the error-correcting properties of the Brownian circuit and probe the transition to an approximate 2-design. 
We conclude by discussing the implications of this work and future directions in section~\ref{sec:discussions}.

\section{Local Brownian circuits}\label{sec:model}

We consider a Brownian circuit on $N$ qubits in a chain, with the Hamiltonian
\bea
H_{t} = \sum_{\langle i,j\rangle}^{N}\sum_{\alpha,\beta} J_{t,ij}^{\alpha\beta} \  \sigma_{i,\alpha}\sigma_{j,\beta},
\eea
where $\alpha, \beta$ label the Pauli indices of the local Pauli matrices $\sigma_{i}$, interacting between nearest neighbor pairs $\langle i,j \rangle$. 
$J_{t,\alpha\beta}$ is a normal random variable uncorrelated in time, defined via the following properties
\bea \label{eq:gaussian}
    &\mathbb{E}\left[J_{t,ij}^{\alpha\beta}\right] = 0 \nn \\
    &\mathbb{E}         \left[J_{t,ij}^{\alpha\beta}J_{t^{\prime},ij}^{\alpha^{\prime}\beta^{\prime}}\right] = \frac{J\delta_{tt^{\prime}}}{\delta t}\delta_{\alpha \alpha^{\prime}}\delta_{\beta \beta^{\prime}}.
\eea
$\mathbb{E}$ denotes the average according to the distribution.

\subsection{Effective Hamiltonian description}

We integrate over the random couplings to get an effective Hamiltonian in the replica space.
To this end, let's first consider a unitary evolution of a density matrix, $\rho' = U \rho U^\dag $.
Explicitly writing out the indices, it is 
\bea 
    \rho'_{a'b'} = \sum_{a,b} U_{a' a} \rho_{ab} U^\dag_{bb'} =  \sum_{a,b} U_{a' a}  U^\ast_{b'b} \rho_{ab}.
\eea
In the second term, we transpose $U^\dag$, and use the fact that $(U^\dag)^T = U^\ast$. 
Viewed as a tensor, the time evolution can be expressed by an operator $U\otimes U^\ast $ acting on a state $\sum_{ab} \rho_{ab} |a\rangle \otimes |b \rangle$. 
This is essentially the Choi–Jamiołkowski isomorphism (the operator-state mapping)~\cite{choi_completely_1975,jamiolkowski}. 
Now we can extend this to two replicas.
Since most of our discussion is focused on two replicas, we derive an effective Hamiltonian for two replicas.
Notice that it is straightforward to generalize the derivation to $k$ replicas and to arbitrary number of qubits on each node~\cite{jian2022linear}.
Because the random couplings at different time are uncorrelated, the central quantity is the instantaneous time evolution (for a small time interval $\delta t$) operator for the four contours,
\bea
    U_1(\delta t) \otimes U_2(\delta t)^\ast \otimes U_3(\delta t) \otimes U_4(\delta t)^\ast,
\eea
where $U_{a}$, $a=1,2,3,4$ denotes the unitary evolution operator generated by the Brownian spin Hamiltonian $U_a(\delta t) = e^{-i \delta t H_{t,a}}$ acting on the four Hilbert spaces. 
It includes two replicas, each of which contains a forward contour $a=1,3$ and a backward contour $a=2,4$.
The complex conjugate is due to the Choi–Jamiołkowski isomorphism, as demonstrated above.

The average over the random coupling reads
\bea
    && \mathbb{E} \left[ U_1(\delta t) \otimes U_2(\delta t)^\ast \otimes U_3(\delta t) \otimes U_4(\delta t)^\ast \right] \\
    &=& \int DJ P[J] \exp\left( \sum_a (- i)^{a} \delta t \sum_{\langle i,j\rangle}\sum_{\alpha,\beta} J_{t,ij}^{\alpha\beta} \tau_{i,a}^{\alpha}\tau_{j,a}^{\beta} \right),  \nn
\eea
where $ \tau_{i,1}^\alpha = \tau_{i,3}^\alpha = \sigma_{i}^\alpha $, and $ \tau_{i,2}^\alpha = \tau_{i,4}^\alpha = (\sigma_{i}^\alpha)^\ast $, $\alpha = 1,2,3$. Here $\sigma^\alpha$ denotes the Pauli matrix, and the complex conjugate for the $a=2,4$ contour is due to the backward evolution. 
$DJ = \prod_{\langle i, j \rangle} \prod_{\alpha, \beta} dJ_{t,ij}^{\alpha\beta} $, and $P[J]$ denotes the Gaussian distribution specified by~(\ref{eq:gaussian}). 
Integrating over the random couplings results in an effective Hamiltonian,
\begin{align}
    && \mathbb{E} \left[ U_1(\delta t) \otimes U_2(\delta t)^\ast \otimes U_3(\delta t) \otimes U_4(\delta t)^\ast \right] = e^{-\delta t H_{\text{eff}}}, \label{eq:imagine}
\end{align}
with 
\bea \label{eq:hamiltonian}
    && H_{\text{eff}} =  \frac{J}2 \sum_{\langle i,j\rangle} \sum_{a,b} (-1)^{a+b} (\vec{\tau}_{i,a} \cdot \vec{\tau}_{i,b}) (\vec{\tau}_{j,a} \cdot \vec{\tau}_{j,b}). \label{eq:hamiltonian}
\eea
This Hamiltonian describes a spin chain with four spins per site, denoted by $\vec \tau_{i,a}$, $a=1,2,3,4$.
In the following, we will see that this Hamiltonian can describe various information phases and phase transitions, such as dynamical computational transition, error correcting transition, etc. 
Similarly, for a finite time evolution, we  have 
\bea
    \mathbb{U} \equiv \mathbb{E}\left[ U_{t} \otimes U_{t}^\ast \otimes U_{t} \otimes U_{t}^\ast\right] = e^{-H_\text{eff} t}.
\eea
Here we use $U_t = e^{-\int dt H_t} $ to denote the unitary generated by the Brownian circuit for a time interval $t$.

%An important comment is that Eq.~\ref{eq:imagine} is true for any $\delta t$, and no small $\delta t$ limit is considered.

\subsubsection*{Numerical Implementation}

We simulate imaginary time evolution in the replica Hilbert space using the TEBD algorithm. The local Hilbert space is $\mathbb{C}_{2}^{\otimes 4}$, reflecting the two replicas and two time contours per replica. To simulate $\exp\left(- t H_{\text{eff}}\right)$ we now need to Trotterise the TEBD evolution with $\Delta t$ as the time step, we take the energy-scale $J = 1/\Delta t$. This ensures that $\Delta t \cdot H_{\text{eff}}$ is dimensionless with the energy scale set to $1$, and with the evolved time $t$ as non-negative integers. All calculations are performed using the TeNPy Library~\cite{tenpy}.

\subsection{Replica permutation symmetry}\label{sec:symmetries}

The Hamiltonian~Eq.~\ref{eq:hamiltonian} is invariant under replica re-labelings, and has the symmetry group,
\bea
    \left(S_{2}\times S_{2}\right)\rtimes \mathbb{Z}_{2},
\eea
where $S_{2} \times S_2$ is the permutation group on the two replica labels. 
The outer $\mathbb{Z}_{2}$ arises from the symmetry of shuffling between the two time-conjugated copies after taking the complex conjugation~\footnote{Note that since the variance of coupling is independent of $\alpha = x,y,z$, the resulted Hamiltonian also enjoys a $SU(2)$ symmetry for each site. But our results do not rely on this symmetry.}.
Put simply, each of the $S_2$ transformation swaps $\tau_{i,1}^\alpha \leftrightarrow \tau_{i,3}^\alpha$ or $\tau_{i,2}^\alpha \leftrightarrow \tau_{i,4}^\alpha$, whereas the $\mathbb{Z}_2$ exchanges $\tau_{i,1}^\alpha \leftrightarrow \tau_{i,2}^\alpha$ and $\tau_{i,3}^\alpha \leftrightarrow \tau_{i,4}^\alpha$ simultaneously.

It is easy to see that the Hamiltonian can be brought into a sum of squares,
\bea
    H_{\text{eff}} = \frac{J}{2} \sum_{\langle i, j \rangle }\sum_{\alpha,\beta} \left( \sum_a (-1)^a \tau_{i,a}^\alpha \tau_{j,a}^\beta \right)^2.
\eea
Therefore, the eigenvalues are no less than zero.
Two ground states are $|\text{id}\rangle\rangle^{\otimes N}$ and $|\text{swap}\rangle\rangle^{\otimes N}$, where
\bea
    |\text{id}\rangle\rangle &=& 
    \frac12(|0000 \rangle + |0011 \rangle + |1100 \rangle + |1111 \rangle ), \nn \\
   |\text{swap}\rangle\rangle &=& 
   \frac12(|0000 \rangle + |1001 \rangle + |0110 \rangle + |1111 \rangle ).
\eea
Here, we use $\left|0\right>$ and $\left|1\right>$ to denote $\pm$ eigenstates of the $\sigma_{z}$ Pauli operator. The name of the state indicates that $\left| \text{id} \right> \rangle$ is a product of an EPR state of the first and second spins and an EPR state of the third and fourth spins, and $\left| \text{swap} \right> \rangle$ is a product of an EPR state of the first and fourth spins and an EPR state of the second and third spins. 
Using the properties of EPR pairs, namely, $\tau_{1}\kettid =\tau_{2}\kettid$, $\tau_{3}\kettid =\tau_{4}\kettid$, $\tau_{1}\kettswap =\tau_{4}\kettswap$, $\tau_{2}\kettswap =\tau_{3}\kettswap$ we can see that every square in the Hamiltonian vanishes, so that these two states are ground states with zero energy.

The permutation symmetry is spontaneously broken by the ground state, and when the low-energy physics is concerned, our model is essentially equivalent to an Ising model.
Notice that we can organize the permutation transformation such that one of them permutes the second and the fourth spins (we denote this by $S_2^r: \tau_{i,2}^\alpha \leftrightarrow \tau_{i,4}^\alpha$), while the other permutes both the first and the third spins as well as the second and the fourth spins. 
Then $S_2^r$ can transform one ground state to the other, and only $S_2^r$ is spontaneously broken. 

Our model~Eq.~\ref{eq:imagine} transforms the real-time evolution along the four contours into an imaginary-time evolution that progressively projects onto the ground state subspace of the Hamiltonian described in Eq.~\ref{eq:hamiltonian}. 
This imaginary-time evolution allows us to capture the dynamics of several important quantum information quantities. 
One such quantity is the collision probability, which measures the degree of anticoncentration and corresponds in the replica model to the overlap between the time-evolved state and a final state (to be specified later). 
The magnitude of this overlap is determined by the excitation gap present in the Hamiltonian~Eq.~\ref{eq:hamiltonian}. 
In one-dimensional systems, the elementary excitation takes the form of domain walls, which possess a finite energy gap and exhibit logarithmic entropy. 
As a result, the process of anticoncentration requires a timescale proportional to $\log N$, where $N$ represents the system size.

\subsection{Effective Hamiltonian with local noise}

Since we would also like to investigate the effect of quantum noise, we now consider imperfect time evolution due to the presence of quantum errors. 
The unitary time evolution operators are replaced by a quantum channel. A local depolarization channel is given by 
\bea
    \rho \rightarrow (1-\lambda) \rho + \frac\lambda3 \sum_{\alpha = 1,2,3} \sigma_i^\alpha \rho \sigma_i^\alpha,
\eea
where $0 \leq \lambda < 3/4 $ for complete positivity.

%To be concrete, we describe two different quantum channels.
%A dephasing channel with strength $\lambda$ at site $i$ is given by
%\bea
%    \rho \rightarrow (1-\lambda) \rho + \lambda \sigma_i^v \rho \sigma_i^v, \quad \sigma^v = \vec v \cdot \vec \sigma,
%\eea
%where $\vec v$ is a unit vector.

Using the operator-state mapping, this can be mapped to,
\bea
    %\mathcal N_i^v (\lambda) &=& (1-\lambda) I^{\otimes 2} + \lambda \sigma_i^v \otimes (\sigma_i^v)^\ast, \\
    \mathcal N_i^\text{depol}(\lambda) &=& (1-\lambda) I^{\otimes 2} + \frac\lambda3 \sum_{\alpha=1,2,3} \sigma_i^\alpha \otimes (\sigma_i^\alpha)^\ast,
\eea
where $I$ denotes the identity operator.

The noise can induce a transition of random circuit sampling~\cite{morvan2023phase, ware2023sharp}. 
An observable of such a transition is the cross-entropy benchmarking (XEB), which we will describe in detail later. 
For now, let us just mention that XEB contains two distributions: one from a noiseless quantum circuit, and the other from a noisy quantum circuit.
Therefore, we are again concerned with only two replicas.
Without loss of generality, we assume the noisy replica is described by the first two contours $a= 1, 2$ and the noisy replica is described by the last two contours $a= 3, 4$.  
We upgrate the quantum channel into 
\bea \label{eq:channel}
    %\mathcal N_i^v (\lambda) \rightarrow \mathcal N_i^v (\lambda) \otimes I \otimes I,  %&=& (1-\lambda) I^{\otimes 4} + \lambda \sigma_i^v \otimes (\sigma_i^v)^\ast \otimes I \otimes I, \\
    %\\
    \mathcal N_i^\text{depol}(\lambda) \rightarrow \mathcal N_i^\text{depol}(\lambda) \otimes I \otimes I. %&=& (1-\lambda) I^{\otimes 4} + \frac\lambda3 \sum_{\alpha=1,2,3} \sigma_i^\alpha \otimes (\sigma_i^\alpha)^\ast \otimes I \otimes I. \nn 
\eea
The identity operators for the last two Hilbert space is clear since the second replica is noiseless. Thus, the noise occurs for the first replica with the first and second copies of the Hilbert space.

It is not hard to see that the channel can be equivalently described by a perturbation described by the following effective Hamltonian,
\begin{align} \label{eq:perturbation}
    %H_v(\lambda) &=&  \frac1{2\delta t} \log(\frac1{1- 2\lambda})  \sum_i (1 -\tau_{i,1}^v \tau_{i,2}^v), \\
    H_\text{depol}(\lambda) = \frac3{4\delta t} \log(\frac1{1- \frac43\lambda})  \sum_i \left(1 - \frac13 \sum_\alpha \tau_{i,1}^\alpha \tau_{i,2}^\alpha\right). 
\end{align}
%where $\tau^v_{i,a } = \vec v \cdot \vec \tau_{i,a}$. 
Here, we assume the noise occurs at each site with the same strength. Since $0 \leq \lambda < \frac34$ for the depolarizing channel, the prefactor is positive.

Essentially, the perturbation explicitly breaks the permutation symmetry. 
The state $\left| \text{id} \right> \rangle^{\otimes N}$ is still an eigenstate of these two perturbations with eigenvalue zero, whereas, the state $\left| \text{swap} \right> \rangle^{\otimes N}$ obtains a finite positive energy, i.e.,
\begin{align}
    \langle\left< \text{swap} \right|^{\otimes N} H_\text{depol}(\lambda) \left| \text{swap} \right> \rangle^{\otimes N} = \frac{3N}{4\delta t} \log(\frac1{1- \frac43\lambda}). 
\end{align}

Therefore, the presence of noise effectively lifts the degeneracy between the two ground states and biases the system towards the state $\left| \text{id} \right> \rangle^{\otimes N}$.
In the regime of low-energy physics, the noise can be treated as an external field that explicitly breaks the Ising symmetry and favors a particular state. 
For notation simplicity, we denote the local Zeeman energy as $\epsilon$.
For the depolization channel, %$\epsilon_v = \frac{1}{2\delta t} \log(\frac1{1- 2\lambda})$ and 
the effective Zeeman energy is $\epsilon = \frac{3}{4\delta t} \log(\frac1{1- \frac43\lambda})$.
When $\lambda$ is small, we can deduce that $\epsilon \approx \frac{\lambda}{\delta t}  $. 
Note that $\lambda $ is dimensionless and $\delta t$ is the unit of energy.

In the symmetry breaking phase, the external field acts as a relevant perturbation with a scaling dimension of one. 
Consequently, a first-order transition occurs at an infinitesimally small noise strength, which is independent of the system size. 
However, if the noise strength is appropriately scaled down by a factor of $1/N$, which compensates for its relevant scaling dimension, the transition can take place at a finite, size-dependent noise strength given by $\epsilon \sim 1/N$. 
This noise-induced transition also manifests as a first-order phase transition. 
The first-order transition exhibits a finite size scaling, and is distinguished from a second-order transition~\cite{binder1984finite}.
We will confirm it via a systematic finite size scaling.

%\R{[explain relation between mutual purity and the model]}
%Finally, we explore the coding properties of the Brownian circuits. 
%To this end, we consider the capability of encoding a single qubit in the Brownian circuit.
%After the reference qubit scrambles in the Brownian circuit by a time $t$, its error correcting property is quantified

%A summary of various quantum informative transitions is given in Table~\ref{table:summary}.

\section{Anticoncentration and computational hardness of sampling Brownian circuits} \label{sec:anticonc}

\subsection{Anticoncentration}
%\textcolor{red}{Maybe bring some of the intro text into this section}
It is well-known that random circuits generate output states that are anti-concentrated, which roughly means that the probability distribution of the classical bitstrings generated by measuring the output state of a random circuit in the computational basis, is well spread out and not concentrated on a few bit-strings. 
Naturally, this also implies that classical sampling of these bitstrings will be hard. 
Two key ingredients underpin random circuit sampling. Firstly, anticoncentration asserts that the distribution deviates only slightly from a uniform distribution. 
This property is typically required in hardness proofs. However, anticoncentration can be easily attained by applying a Hadamard gate to all qubits. Therefore, we need the second ingredient, randomness, to eradicate any discernible structure in the circuit. Given that randomness is inherent in our model, we are intrigued by whether the distribution exhibits anticoncentration and, if so, at what time (depth) it occurs.
This indicates a transition in computational complexity, wherein the system shifts from a region that is easily achievable by classical means to a region that becomes challenging for classical algorithms.

When the random circuits are generated by a particular ensemble of local quantum gates, a key diagnostic of the complexity of the ensemble is the time it takes to anti-concentrate the output states. 
Concretely, we can compute the collision probability, which is defined as the probability that the measurement outcomes of two independent copies of the random circuit agree with each other, i.e. $\sum_{s}p_{U}(s)^{2}$, where $p_{U}(s) = \left|\bra{s}U\ket{0}\right|^{2}$, for a given bitstring $s$. 
We are interested in the ensemble averaged collision probability which can be readily expressed as transition amplitude in the replicated dynamics,
\begin{align}
    Z &= \mathbb{E}_{J}\sum_{s}p_{U}(s)^{2} = \sum_s \langle \langle s^{\otimes 4} | \mathbb{U} | 0^{\otimes 4} \rangle \rangle, 
\end{align}
where $\mathbb{U} = \mathbb{E}\left[ U_{t} \otimes U_{t}^\ast \otimes U_{t} \otimes U_{t}^\ast\right]$ can be represented by an imaginary time evolution with a replica Hamiltonian defined in Eq.~\ref{eq:hamiltonian}.

We identify the circuit to have reached anti-concentration if $Z \approx c 2^{-N}$ and to not have anti-concentration if $Z\geq e^{N^{c}}2^{-N}$ for some $O(1)$ constant $c$. 

\begin{figure}[h]
    \centering
    \includegraphics[width = \columnwidth]{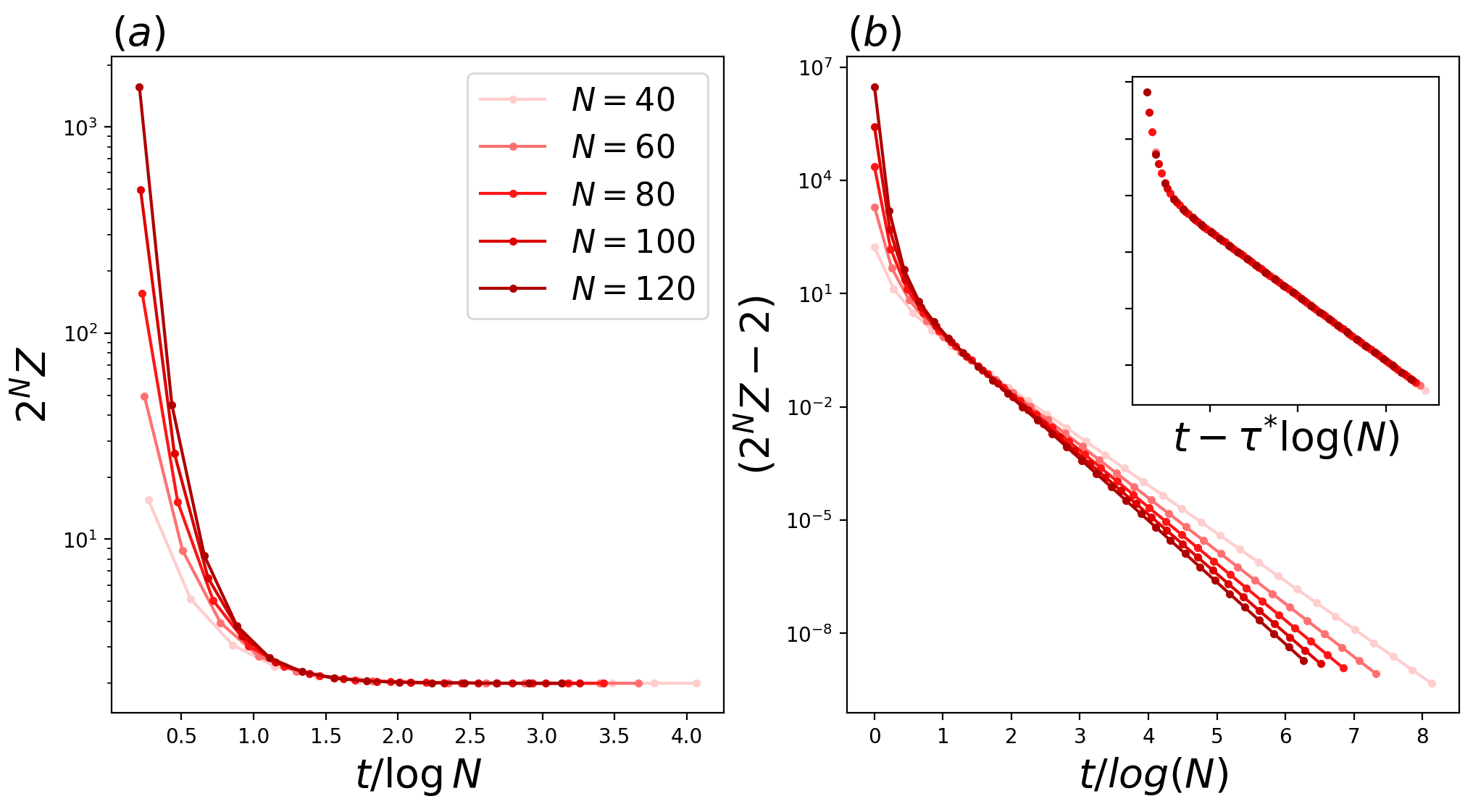}   \caption{\textbf{Anticoncentration in Brownian circuit:} In (a) we show that $Z \sim 2^{-N}$ in $t/\log N = \tau^{*} \approx 1.5$ time. In (b) we show the data collapse, which is consistent with Eq.~\ref{eq:collision_approx}  }
    \label{fig:anticonc_1}
\end{figure}

In Fig.~\ref{fig:anticonc_1} we study the averaged collision probability in a 1d Brownian circuit by the tensor network simulations. 
We find that the Brownian circuit anti-concentrates in $\log{N}$ depth, which is consistent with the fact that local Haar random circuits anti-concentrate in $\Omega(\log N)$ depth in 1d \cite{Dalzell_2022}. 
Furthermore, in Fig.~\ref{fig:anticonc_1}b, we show data collapse which is consistent with the following approximate form for the collision probability,
\bea
2^{N} Z = 2+ c_{1}e^{-c_{2}(t-\tau^{*}\log N)},
\label{eq:collision_approx}
\eea
for some $O(1)$ constants $c_{1}$ and $c_{2}$. This expression can be justified by the effective Hamiltonian picture as follows.

Because $\mathbb U = e^{-H t}$, with $H$ given by Eq.~\ref{eq:hamiltonian}, it effectively projects the initial state $\left|0^{\otimes 4} \right>\rangle$  to the ground state $2^N \mathbb U \left|0^{\otimes4} \right>\rangle \approx \left| \text{id} \right>\rangle^{\otimes N} + \left| \text{swap} \right>\rangle^{\otimes N} + \text{excitations} $. 
The leading contribution of excitations is given by a single domain wall (since we have used open boundary condition, a single domain wall is allowed), $\left| \text{DW}_k \right> \rangle \approx \left| \text{swap} \right>\rangle^{\otimes k} \otimes \left| \text{id} \right>\rangle^{\otimes (N-k)}$, $k=1,...,N-1$.
Therefore, the multiplicity of such an excitation is proportional to $N$.
The excitation energy $\Delta$, on the other hand, is a constant independent of $N$, and it contributes to an exponential function $e^{-\Delta t}$. 
Therefore, according to this picture, the prediction for the collision probability reads
\bea
    2^N Z \approx 2 + N e^{-\Delta t} = 2 + e^{-\Delta (t- \frac1{\Delta} \log N)},
\eea
where we have noticed that $\sum_s \langle \left< s | \text{id} \right> \rangle = \sum_s \langle \left< s | \text{swap} \right> \rangle =\sum_s \langle \left< s | \text{DW}_k \right> \rangle = 1$.
This result is consistent with the data collapse. 
In particular, it is clear that the transition time $\log N$ is due to the entropy of the domain wall excitation.

\subsection{Hardness of classical simulation}\label{sec:anticonc_hard}

As a consequence of anticoncentration and randomness, classical simulation of the output probabilities of the Brownian circuits after $\log N$ depth is expected to be hard. 
In this section, we show that, with respect to a particular algorithm for approximate classical simulation, there is a computational hardness transition at $t\sim \log N$ depth.

We study the computational hardness of the Patching algorithm introduced in~\cite{Napp_2022,brandao2019finite}. 
Heuristically, the algorithm attempts to sample from the marginal probability distribution of spatially separated patches, and then combine the results together. 
This succeeds in $poly(N)$ time if the output distribution of the state generated by the circuit has decaying long-range correlations. 
Without going into the details of the algorithm itself, we study the condition on the long-range correlations for which the algorithm is expected to successfully sample from the output distribution. 

Consider a tripartition of $N$ qubits into $A\cup B\cup C$, such that $dist(A,C)\geq l $. 
For the output probability distribution $p_{U}(s) = \left|\bra{s}U\ket{0}\right|^{2}$, we consider the conditional mutual information between the regions $A$ and $C$ conditioned on $B$, as in $I(A:C|B)_{p} = S(AB)_{p}+S(BC)_{p}-S(B)_{p}-S(ABC)_{p}$, where the $S(A)$ refers to the entropy of the marginal distribution of $p$ on the region $A$. 
The output distribution is defined to have $f(l)-$ Markov property if $I(A:C|B)_{p} \leq f(l)$.

We quote the main Theorem about the condition for successful Patching algorithm from~\cite{Napp_2022}:

\begin{theorem}
    Patching algorithm succeeds in $poly(N)$ time to sample from a probability distribution arbitrarily close in total variation distance to the exact output distribution $p_{U}(s)$ of a quantum circuit on $N$ qubits, if $p_{U}(s)$ has $e^{-\Omega(l)} $ Markov property, for a suitable choice of the length-scale parameter $l$.
\end{theorem}

\begin{figure}
    \centering
    \includegraphics[width = \columnwidth]{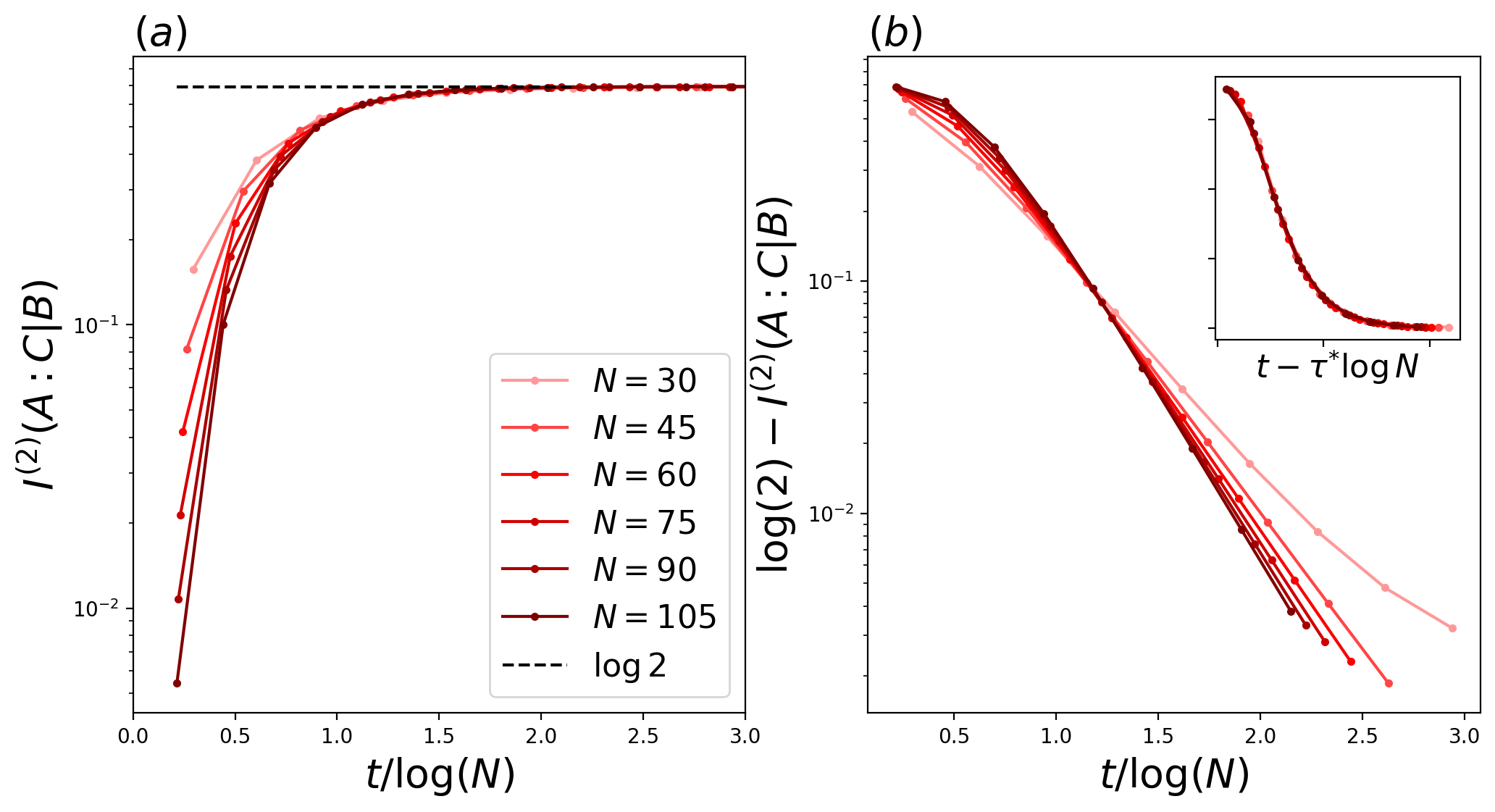}
    \caption{\textbf{Phase transition in the R\'enyi-2 CMI of $p_{U}(s)$:} Consider an equal tripartition of the spin chain such that $|A| = |B| = |C| = N/3$, with $B$ separating the regions $A$ and $C$, and we plot R\'enyi-2 CMI $I^{(2)}(A:C|B)$ in (a) and (b). As $N$ is increased, for $t<\tau^{*}\log N$ ($\tau^{*}\approx 1.2$), the CMI decays with $N$ as $\sim e^{-N}$. For larger times $t>\tau^{*}\log N$ the CMI doesn't decay with $N$ and asymptotes to $\log 2$.}
    \label{fig:cpt}
\end{figure}

In the local Brownian circuits introduced earlier, we can directly compute the averaged R\'enyi-2 version of the conditional mutual information (CMI) of the output distribution $p_{U}(x)$, i.e. $I^{(2)}(A:C|B)_{p} = S^{(2)}(AB)_{p}+S^{(2)}(BC)_{p}-S^{(2)}(B)_{p}-S^{(2)}(ABC)_{p}$, as a function of time $t$. 
In Fig.~\ref{fig:cpt}a and b, we study the R\'enyi-2 CMI for an equal tripartition of the qubit chain (i.e. $|A| = |B| = |C| = N/3$), and find that there is a transition at $\log N$ depth. 
In particular, at long times, $I^{(2)}(A:C|B)_{p}$ asymptotes to $\log 2$, indicating long-range correlations in the output probability distribution. 
At short times, the data is consistent with $I^{(2)}(A:C|B)_{p}\sim O(e^{-N})$. There is, furthermore, a sharp transition at $t\sim \tau^{*}\log N$ at $\tau^{*}\approx 1.2$. 
The data collapses as a function of $t-\tau^{*}\log N$ as shown in Fig.~\ref{fig:cpt}b inset, indicating the same statistical mechanical interpretation as the collision probability. 
Even though this result is for the R\'enyi-2 version of the CMI, and not the actual CMI itself, it provides evidence that the anticoncentration transition corresponds to an actual phase transition in computational hardness of classical estimation of the output probabilities of the random circuit.

\section{Noisy Brownian circuits}\label{sec:noise}

Random circuit sampling is widely implemented in experiments to show quantum advantage.
However, sufficiently large noise can diminish the quantum advantage. 
It was reported recently a noise induced phase transition in random circuit sampling~\cite{morvan2023phase,ware2023sharp,dalzell2021random, Aharonov_2023}. 
For weak noise, the cross-entropy benchmarking provides a reliable estimate of fidelity.
Whereas, for strong noise, it fails to accurately reflect fidelity.

\subsection{Cross-entropy benchmarking} 

In the random circuit sampling, we start from a product state $\rho_0 = \left|0\right>^{\otimes N}  \left< 0\right|^{\otimes N}$ (The initial state does not really matter, and we choose this just for simplicity), and evolve the state using the Brownian spin Hamiltonian.
For brevity, we denote the unitary generated by the Brownian spin model as $U$.
In an ideal case, i.e., there is no noise, the final state is 
\bea
    \rho = U \rho_0 U^\dag.
\eea
A measurement is performed on the computational basis, and this will generate a probability distribution,
\bea
    p(s) = \langle s | \rho | s\rangle,
\eea
where $s$ denotes the bit string. 

In a real experiment, the implementation of Brownian spin Hamiltonian is not ideal because errors can occur. 
In this case, the time evolution of the system is, in general, not unitary and should be described by a quantum channel,
\bea
    \rho_\text{err} = \mathcal N (\rho_0).
\eea
Here, $\mathcal N$ denotes the noise channel.
The probability distribution for a bit string $s$ is now given by
\bea
    q(s) = \langle s | \rho_\text{err} | s \rangle.
\eea

We are interested in the cross entropy benchmarking (XEB), defined as follows,
\bea
    \chi_\text{XEB} = 2^N \sum_s p(s) q(s) - 1,
\eea
where $p(s)$ is an ideal distribution (which in practice can be estimated by classical simulations), and $q(s)$ is the probability distribution sampled from real experiments.
Since the circuit involves Brownian variables, we consider the average over these random variables, $\mathbb E(\chi_\text{XEB})$.

\subsection{XEB in the replica model} 

Using the operator-state mapping, 
\bea
    \sum_s q(s) p(s) = \sum_s \langle \langle s | \mathcal N \otimes U \otimes U^\ast |0 \rangle \rangle,
\eea
where $U$ is the unitary generated by the Brownian spin model, and $\mathcal N$ denotes the channel generated by both the Brownian spin model and the errors.
For simplicity, we will denote $\mathbb{E}[\mathcal N \otimes U \otimes U^\ast]= \mathbb U_\text{err}$.
And the initial and final states are the same as in the collision probability.
Actually, the collision probability is closely related to the noiseless XEB.
%\bea
%    |0 \rangle \rangle = |0 \rangle^{\otimes 4 L}, \quad |s\rangle \rangle = |s\rangle^{\otimes 4}.
%\eea
%Notice that the noisy channel cannot be written as a tensor product form. 

%We first discuss the noiseless case. 
%The average over Brownian variables leads to an effective Hamiltonian~Eq.~\ref{eq:hamiltonian}, i.e.,
%\bea
%     \sum_s \langle \langle s | \mathbb{E}\left[ U \otimes U^\ast \otimes U \otimes U^\ast\right] | 0 \rangle \rangle  = \sum_s \langle \langle s | e^{-H T} | 0 \rangle \rangle,
%\eea
%where $T$ is the total evolution time. 
%The average XEB becomes a transition amplitude between two states $|0\rangle \rangle$ and $| s \rangle \rangle$.

Consider imperfect time evolution due to the presence of quantum errors. 
To this end, after integrating over the Brownian variable, we arrived at the imaginary-time evolution given by
\bea
     \sum_s \langle \langle s | \mathbb{U}_\text{err} | 0 \rangle \rangle  = \sum_s \langle \langle s | e^{-(H+H'(\lambda)) t} | 0 \rangle \rangle,
\eea
where $H'(\lambda)$ is the perturbation caused by the noise. 
The example of dephasing and depolarizing channels are given by~Eq.~\ref{eq:perturbation}. 
The average XEB then reads 
\bea
    \mathbb{E} [\chi_\text{XEB}] = 2^N \sum_s \langle \langle s | e^{-(H+H'(\lambda)) t} | 0 \rangle \rangle - 1,
\eea
On the other hand, the average fidelity is given by 
\bea
    \mathbb E [F] &=& 2^N \langle \langle \text{swap} |^{\otimes N} \mathbb{E} [\mathcal N \otimes U \otimes U^\ast] | 0^{\otimes4} \rangle \rangle \\
    &=& 2^N \langle \langle \text{swap} |^{\otimes N} e^{-(H+H'(\lambda)) t} | 0^{\otimes4} \rangle \rangle.
\eea
Comparing it with the XEB, we can see that the difference comes from the final state.

As discussed before, the noise lifts the degeneracy and behaves as an external field.
We denote the local Zeeman energy by $\epsilon$. 
The Zeeman field is a relevant perturbation even in the symmetry breaking phase.
We will show that the competition between one of the lifted ground state and the excited state leads to a first-order transition at a finite noise rate $\epsilon N \sim $ const. 
We will also perform a finite size scaling analysis to verify such a first-order transition in the following.

%Here, we describe two different quantum channels.
%A dephasing channel with strength $\lambda$ at site $i$ is given by
%\bea
%    \rho \rightarrow (1-\lambda) \rho + \lambda \sigma_i^v \rho \sigma_i^v, \quad \sigma^v = \vec v \cdot \vec \sigma,
%\eea
%where $\vec v$ is a unit vector. 
%A depolarization channel is given by 
%\bea
%    \rho \rightarrow (1-\lambda) \rho + \frac\lambda3 \sum_{\alpha = 1,2,3} \sigma_i^\alpha \rho \sigma_i^\alpha.
%\eea
%Using the operator-state mapping, the two channels are mapped, respectively, into
%\bea
%    \mathcal N_i^v (\lambda) &=& (1-\lambda) I^{\otimes 4} + \lambda \sigma_i^v \otimes (\sigma_i^v)^\ast \otimes I \otimes I, \\
%    \mathcal N_i^\text{depol}(\lambda) &=& (1-\lambda) I^{\otimes 4} + \frac\lambda3 \sum_{\alpha=1,2,3} \sigma_i^\alpha \otimes (\sigma_i^\alpha)^\ast \otimes I \otimes I,
%\eea

%The noisy channel explicitly breaks the symmetry between $|\text{id} \rangle \rangle $ and $|\text{swap} \rangle \rangle $ as we can simply notice 
%\bea
%    \langle \langle \text{id} | \mathcal N_i^v(\lambda) | \text{id} \rangle\rangle &=& 1, \\
%    \langle \langle \text{swap} | \mathcal N_i^v(\lambda) | \text{swap} \rangle\rangle &=& 1-\lambda.
%\eea

%The quantum error is modeled by setting the noise quantum channel occurring at random positions. 

We consider the evolution of XEB as a function of time.
In the long-time limit, we expect the time-evolved state is a superposition of the ground state with a few low-lying excitations.
It can be approximately written as
\bea
    && 2^N \mathbb{U}_\text{err} \left| 0^{\otimes 4} \right>\rangle \approx  \left|\text{id} \right> \rangle^{\otimes N} + e^{-N \epsilon t} \left| \text{swap} \right>\rangle^{\otimes N} \\
    && + e^{-\Delta t} \sum_k e^{ -k\epsilon t} \left|\text{DW}_k \right>\rangle + \sum_k e^{-(2\Delta + \epsilon) t} \left|\text{SF}_k \right> \rangle \nn
\eea
where $\Delta $ is the local energy cost of a domain wall, and $\epsilon$ are the local energy cost and the Zeeman energy of a local spin flip.
We have included both domain wall excitations and local spin flips, $\left|\text{SF}_k \right>\rangle = \left|\text{id} \right> \rangle^{\otimes k-1} \otimes \left|\text{swap} \right> \rangle \otimes \left|\text{id} \right> \rangle^{\otimes N-k} $. 
Note that the domain wall excitation can lead to an extensive energy cost, but we need to include them because the external field scales $\epsilon \sim 1/N$. 
Therefore, the average XEB at late time is
\bea
    \mathbb E[\chi_\text{XEB}] &=& e^{-N \epsilon t} + e^{-\Delta t} \sum_{k=1}^{N-1} e^{-k \epsilon t} + N e^{-(2\Delta + \epsilon) t}, \nn \\
\eea
%where we have used $2^N \sum_s \langle \left< s |\text{id} \right> \rangle^{\otimes N} = 1$, $2^N \sum_s \langle \left< s |\text{swap} \right> \rangle^{\otimes N} = 1$, and $2^N \sum_s \langle \left< s |\text{SF}_k \right> \rangle = 1$.
On the other hand, the average fidelity is
\bea
    \mathbb E[F] = e^{-N \epsilon t}.
\eea
Actually, the fidelity is lower bounded by $2^{-N}$.
This is because $\left| \text{id} \right> \rangle^{\otimes N}$ and $\left| \text{swap} \right> \rangle^{\otimes N}$ are orthogonal only at the thermodynamic limit $N \rightarrow \infty$.
For a finite $N$, their overlap is $(\langle\left< \text{id} \right|^{\otimes N}) ( \left| \text{swap} \right> \rangle^{\otimes N}) = 2^{-N}$.

It is clear that for the XEB to well estimate the fidelity, we require $e^{-N \epsilon t} \gg  e^{-\Delta t}$. 
If we consider the ratio between them 
\bea
    \frac{\mathbb E [F]}{\mathbb E [\chi_\text{XEB}]} \approx \frac{1}{1+ e^{-\Delta t  + N\epsilon t }}. 
\eea
To the leading order in $N$, there is a noise-induced phase transition at 
$\epsilon_c = \frac{\Delta}{N}$, separating between a weak noise phase, where the XEB well estimates the fidelity, and a strong noise phase, where they do not match.
This is consistent with the scaling dimension analysis.

\subsection{Noise-induced transition}

\begin{figure}
    \centering
    \includegraphics[width = 0.6\columnwidth]{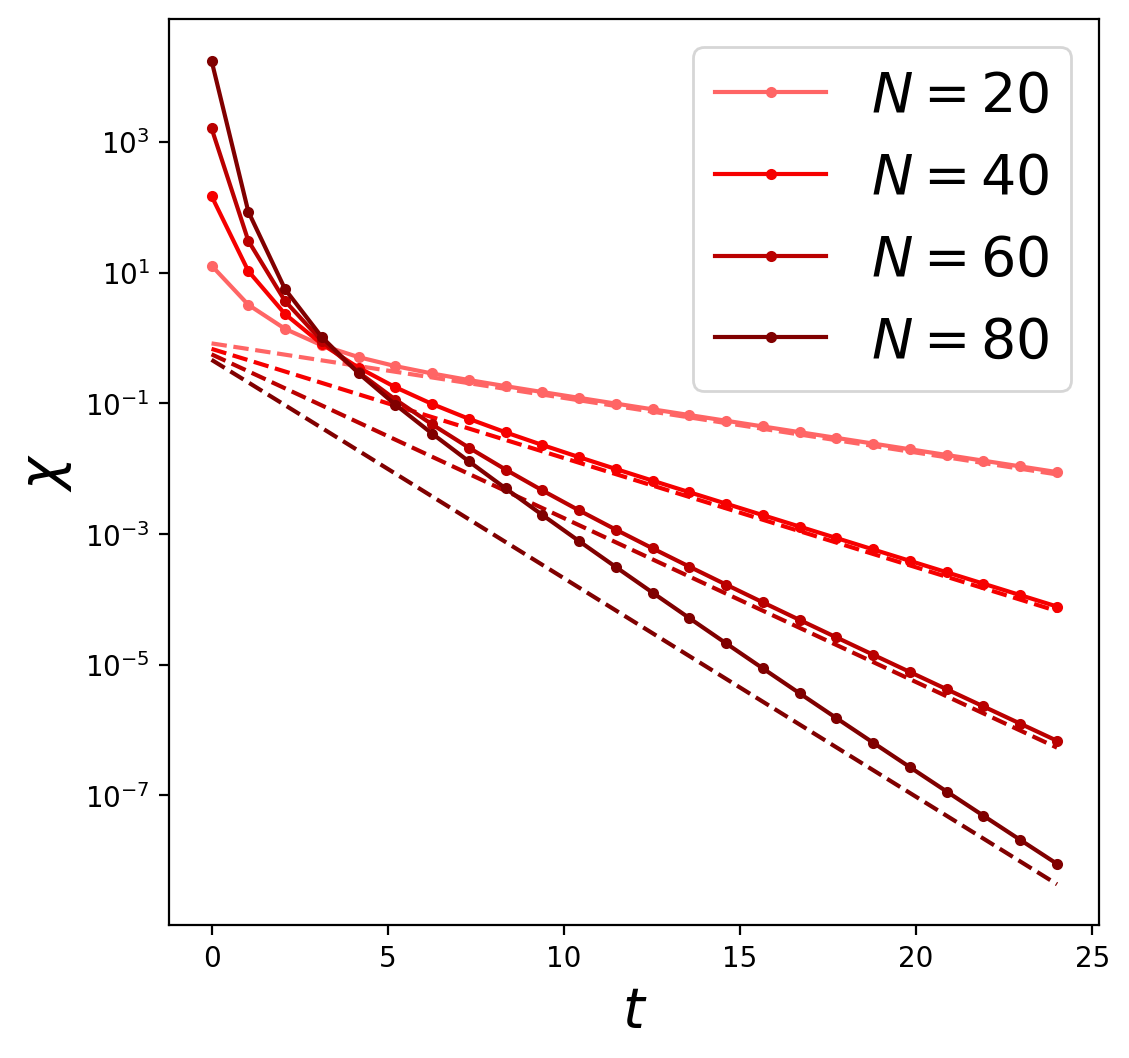}
    \caption{In the presence of noise, there is a depth driven crossover at $t^{*} \sim O(1)$; for $t< t^{*}$ the linear XEB $\chi \sim e^{N}$ and for $t>t^{*}$ $\chi \sim e^{-N\epsilon t}$. We consider a constant noise strength of $\lambda = 0.01$.}% In (b) we find that $\chi - F$, where $F$ is the fidelity of the circuit, has a sharp phase transition, with a scaling collapse shown in the inset.}
    \label{fig:depth_xeb}
\end{figure}

In the short-time region, all kinds of excitations contribution to the XEB, and its evolution is non-universal. 
A crude estimate of the XEB is given as follows,
\bea
    \mathbb{E} [\chi_\text{XEB}] \approx (1 + e^{-(2\Delta + \epsilon) t} )^{N} - 1 .
\eea
This estimate comes from the superposition of all possible spin flips at each site~\footnote{For a more accurate estimate, we need to rescale $N$ by a factor $c_3 < 1$. This is because spin flips do not interact with each other only when they are dilute enough. }.
Here $\Delta$ is the effective local energy cost of a spin flip.
The XEB is exponential in the system size $\sim \exp[N e^{-(2\Delta + \epsilon) t}]$, but this behavior decays exponentially fast. 
Then the XEB will transition to the late-time behavior.
In the long-time limit, since we are at the weak noise phase, we expect the XEB matches fidelity. 
To verify this, We plot the time evolution of XEB (solid curves) and fidelty (dahsed curves) in Fig.~\ref{fig:depth_xeb} for a fixed noise rate.
At the long-time limit, their evolution follows closely.
The fact that the deviation is larger for a bigger $N$ is because we have fixed $\epsilon$. 
It is also clear that the XEB curves exhibit a crossover from a short-time non-universal region to a long-time universal region.

\begin{figure}
    \centering
    \includegraphics[width = \columnwidth]{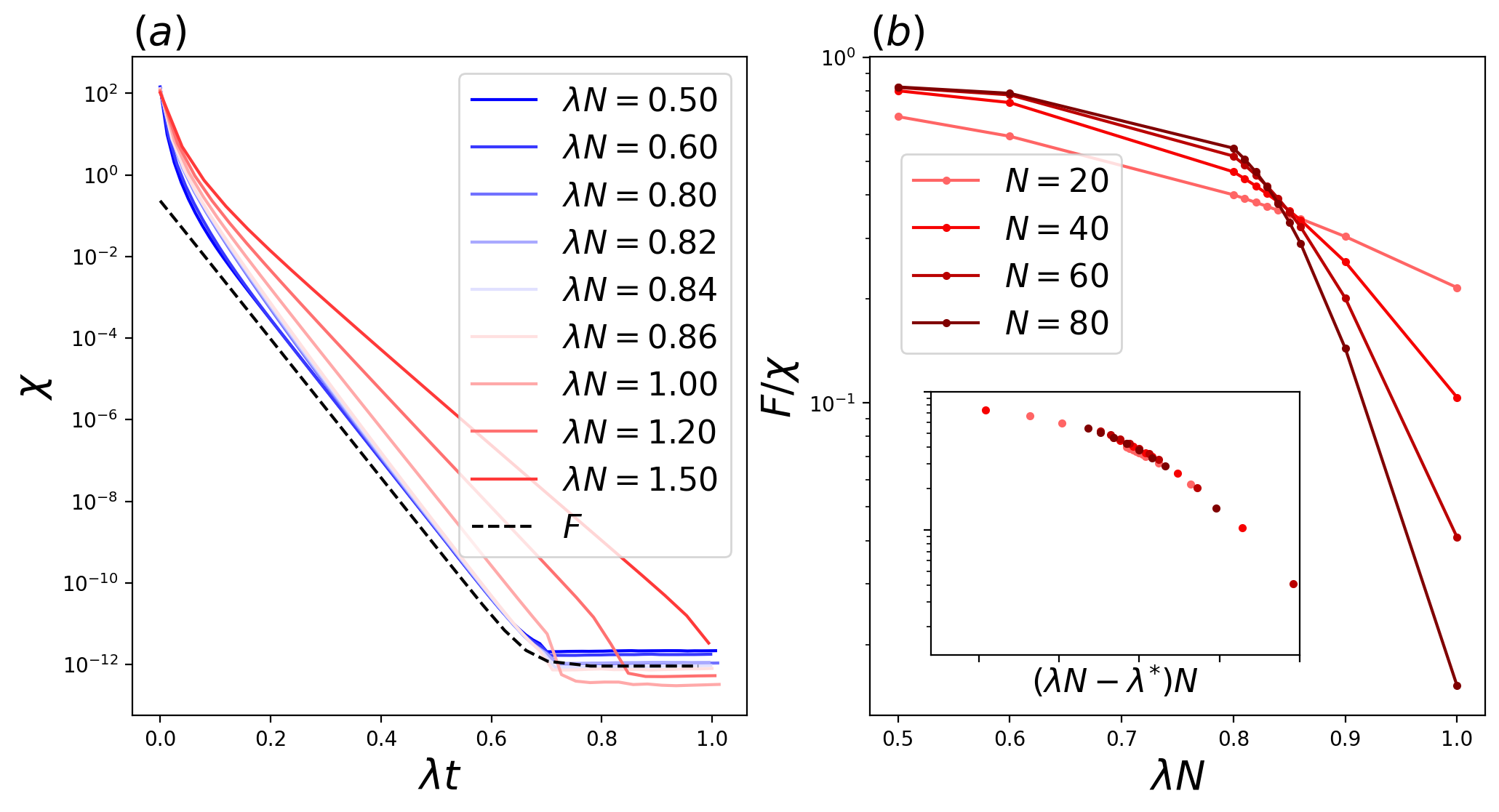}
    \caption{\textbf{Noise driven transition for scaled noise $\lambda \sim c/N$} In (a) we plot $\chi$ for $N = 40$ and for different values of $\lambda N$. We find that for $\lambda N \leq \lambda^{*}N \approx 0.84$ $\chi$ approximates the Fidelity $F$ after an initial fast decay. For $\lambda N >\lambda^{*}N$, however, $\chi$ behaves differently from $F$, which results in a sharp phase transition as shown in (b) in the order parameter $F/\chi$, evaluated at a time $t > t^{*} = O(1)$ of initial decay. The scaling collapse is shown in the inset.}
    \label{fig:noise_xeb}
\end{figure}

In order to show the noise-induced phase transition in our replica model, we plot the time-evolution of the XEB for different noise rates in Fig.~\ref{fig:noise_xeb}a. 
It is clear that when the noise rate is less than $\lambda^\ast \approx 0.84/N $, the XEB tracks the fidelity very well. 
Here the fidelity is shown by a dashed curve. 
Note also that the fidelity has a lower bound given by $2^{-N}$. 

Next, to connect this to the statistical mechanical model and implement a finite size scaling analysis, we consider scaling $t \sim N$ to feature an equal space-time scaling.
The ratio between the fidelity and XEB is plotted in Fig.~\ref{fig:noise_xeb}b for different system sizes. 
The crossing indicates a transition at $\lambda^\ast N \approx 0.84$.
The inset shows data collapse for different sizes as a function of $(\lambda - \lambda^\ast) N^2$, which shows $1/\nu = 2$.

To understand this exponent, we briefly review the finite size scaling at first-order phase transitions.
The finite size scaling near a first-order phase transition is studied in Ref.~\cite{binder1984finite}. 
We briefly repeat the argument here.
In a classical Ising model in $d$ dimensional cube with size $L^d$, the probability distribution of the magnetization $P_L(s)$ in the ferromagnetic phase can be well approximated by a double Gaussian distribution 
\bea
    P_L(s) \propto e^{-(s-M)^2 L^d/\chi} + e^{-(s+M)^2 L^d/\chi},
\eea
here $\chi $ denotes the susceptibility, and $M$ is the average magnetization. 
To incorporate the external field, notice that the probability distribution can be expressed as $P_L(s) \propto e^{-f L^d}$, where $f$ is the free energy density.
From the Ising transition, the free energy is given by
\bea
    f &=& f_0 + \frac{r}2 s^2 + \frac{u}4 s^4 - sH \\
    &=& f'_0 + \frac{u}4(s^2-M^2)^2 - sH,
\eea
where $H$ denotes the external field, $M = \sqrt{-\frac{r}{u}}$ is the average magnetization when $r<0$, and $f_0, f'_0$ are unimportant constants. 
If we approximate the magnetization around $\pm M$, then the double Gaussian distribution reads
\bea
    P_L(s) \propto e^{- ((s-M)^2 - s \chi H)L^d/\chi } + e^{- ((s+M)^2 - s \chi H)L^d/\chi },
\eea
where $\chi = -r$.
It is clear that the distribution will be shifted, and the one near $s = M$ will be amplified.
This probability distribution can serve as a starting point for finite size scaling analysis.
The external field is equipped with scaling dimension $L^{-d}$, implying $\nu = 1/d$. 
Now in our analysis, the Hamiltonian~Eq.~\ref{eq:hamiltonian} corresponds to a 1d quantum system or a 2d classical Ising model, which leads to $\nu = 1/2$, consistent with our scaling data collapse in Fig.~\ref{fig:noise_xeb}b.


\subsection{Hardness of simulating noisy Brownian circuits}

\begin{figure}
    \centering
    \includegraphics[width = \columnwidth]{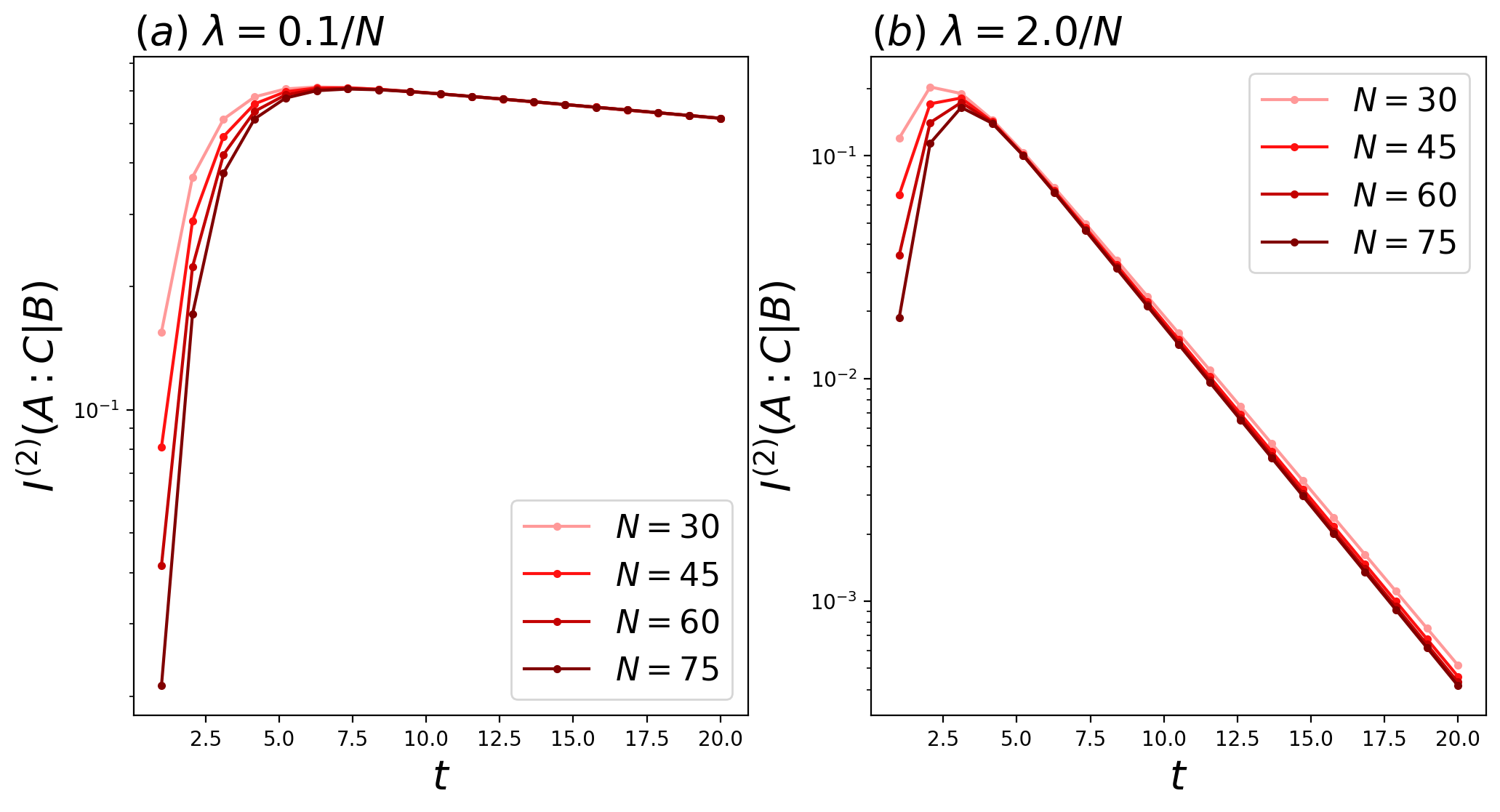}
    \caption{Dynamics of the R\'enyi-2 CMI of the classical output distribution of the noisy Brownian circuit for two noise levels, $\lambda = 0.1/N$ and $\lambda = 2.0/N$ respectively.}
    \label{fig:noisy-hard}
\end{figure}

As we have described, the linear cross-entropy benchmark can be described in the 2-replica formalism, where the noise acts on only one of the replicas. In this section we briefly comment on the hardness of classical simulation of noisy Brownian circuits, by analysing the R\'enyi-2 conditional mutual information of the output distribution $p(s)$ of the noisy circuit, as in Sec.~\ref{sec:anticonc_hard}. In this formulation the noise acts on both replicas. In Fig.~\ref{fig:noisy-hard} we plot the R\'enyi-2 CMI as a function of time for two instances of weak and strong scaled local depolarization channels, with strength $\lambda = \mu/N$ with $\mu = 0.1, 2.0$ respectively. The plots show that the CMI doesn't asymptote to $\log 2$ as the noise-free case, and ultimately decays as $e^{-\mu t}$ without any signature of crossing. This suggests that in the long-time limit, even in the presence of scaled noise, the output distribution remains efficiently estimable using the Patching algorithm~\cite{Napp_2022}.

These numerical results provide evidence that the noise-induced phase transition in the linear cross-entropy benchmark does not signal a phase transition in the hardness of classical simulability of the output distribution of the noisy random circuits. In fact, in the presence of noise, $1+1$d random circuits remain efficiently simulable by the Patching algorithm.

\section{Quantum error correcting codes from Brownian circuits} \label{sec:qecc}

Random circuits scramble local information into global correlations of a state, in a way which is inaccessible to local probes. 
As a result of this, the encoded information can be protected from local noise, thereby leading to the notion of random circuits generating quantum error correcting codes~\cite{Brown_2013,Brown_2015, Gullans_QuantumCoding_2021}.

\subsection{Decoupling by Random circuits}

\begin{figure}[h]
    \centering
    \includegraphics[width = 0.4\columnwidth]{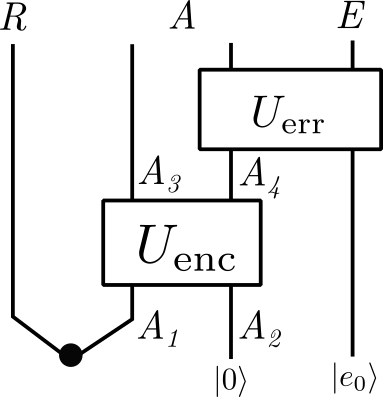}
    \caption{Unitary encoding $U_{\text{enc}}$ of reference $R$ into system $A$. Any noise in $A$ after the encoding can be represented by a unitary operation $U_{\text{err}}$ coupling the part of $A$ where the error acted ($A_{4}$ in the figure), with an environment $E$.}
    \label{fig:qecc_decoupling}
\end{figure}

The intuition as to why random circuits are able to dynamically generate a quantum error correcting code comes from the decoupling principle. 
Consider the setup in Fig.~\ref{fig:qecc_decoupling}, where initial quantum information is initialized in the entangled state between reference $R$ (code subspace) and part of the system $A_{1}\subset A$, such that the dimensions match, $|R| = |A_{1}|$. 
Now $A$ is subjected to an encoding through the random circuit $U_{\text{enc}}$. 
Suppose a part of the system $A_{4}\subset A$ is subjected to a noise channel $\mathcal{N}$. 
By Stinespring dilation, the noise channel can be identified as a unitary coupling with an environment $E$, as shown in Fig.~\ref{fig:qecc_decoupling}. 
If $U_{\text{enc}}$ forms an approximate 2-design, the circuit is able to decouple effectively~\cite{Abeyesinghe_2009}, i.e., the environment $E$ has bounded access to the information encoded in $R$.

Concretely, let us consider local qubit degrees of freedom, such that the Hilbert space dimension of any set $A$ is $d_{A} = 2^{|A|}$. 
Consider the isometric encoding $V:\mathcal{H}_{R} \to \mathcal{H}_{A}$ generated by the circuit $U_{\text{enc}}$, which transforms the basis vectors as follows,
\bea
\ket{\phi_{i}}_{A} \equiv V\ket{i}_{R} = U_{\text{enc}}\ket{i}_{A_{1}}\ket{0}_{A_{2}}.
\eea
Any density matrix $\rho_{R}$ of R is encoded as $V\rho_{R}V^{\dagger}$. 
Suppose the encoded state is now subjected to noise, resulting in the density matrix $\rho_{\text{err}} = \mathcal{N}\left(V\rho_{R}V^{\dagger}\right)$. 
A convenient probe is the noise-affected encoding of a maximally entangled state between the code subspace $R$ and $A_{1}$. By introducing an auxiliary environment $E$ the effect of the noise channel can be represented by a unitary on the combined system and environment, $A \cup E$,
\bea
\ket{\Psi^{\prime}} =  \frac{1}{\sqrt{d_{R}}}\sum_{i = 1}^{d_{R}}\ket{i}_{R} U_{\text{err}} \left(\ket{\phi_{i}}_{A}\ket{e_{0}}_{E}\right).
\label{eq:noisy_state}
\eea
Here, $d_{R}$ refers to the Hilbert space dimension for $R$; if the local degrees of freedom are $q$ dimensional qudits, then $d_{R} = q^{|R|}$. 

By the decoupling theorem, for $U_{\text{enc}}$ which are approximate 2 designs and small enough error, we have a factorized reduced density matrix on $R\cup E$, $\rho_{RE}^{\Psi^{\prime}} \approx \rho_{R}^{\Psi^{\prime}} \otimes \rho_{E}^{\Psi^{\prime}}$. 
The time required by random circuits with locality to approximately form a 2 design is upper bounded by $O(N^{1/d})$ in $d$ dimensions~\cite{Brandao_2016_local, Harrow2023approximate}. 
A probe of the extent of decoupling is the mutual information~\cite{Schumacher_1996}, $I_{\Psi^{\prime}}(R:E) = S(\rho^{\prime}_{R})+S(\rho^{\prime}_{E})-S(\rho^{\prime}_{RE})$.

A central theorem in quantum error correction is the existence of an optimal recovery channel $\mathcal{R}$ that undoes the effect of noise $\mathcal{R}\left(\rho_{\text{err}}\right) = \rho_{R}$, if perfect decoupling has occurred, i.e. $I_{\Psi^{\prime}}(R:E) = 0$ ~\cite{Schumacher_1996}. 
This can be generalized to approximate error correction in the presence of approximate decoupling~\cite{schumacher2001approximate,Beny_General_2010}. 
In particular, the trace distance between the recovered state by a near-optimal recovery channel $\mathcal{R}$, and any encoded state can be bounded by the mutual information computed for $\Psi^{\prime}$,
\bea
\bigg|\bigg|\mathcal{R}\left(\rho_{\text{err}}\right)-\rho_{R}\bigg|\bigg|_{1} \leq \left(I_{\Psi^{\prime}}(R:E)\right)^{1/4}.
\label{eq:approx_qec1}
\eea

\subsection{Approximate error correction in Brownian circuits}

Recently~\cite{balasubramanian2023quantum} derived a similar bound as Eq.~\ref{eq:approx_qec1}, with the right-hand side replaced by a different entropic quantity rather than the mutual information. 
Mutual information is difficult to analytically study because of the associated replica limit in the definition of the von Neumann entropy. 
They instead introduce the \textit{mutual purity} of the noise-affected state in Eq.~\ref{eq:noisy_state}, which is defined as,
\bea
\mathcal{F}_{\Psi^{\prime}}(R:E) = \Tr \left(\rho_{RE}^{\prime 2} - \rho_{R}^{\prime 2}\otimes \rho_{E}^{\prime 2}\right).
\eea

They showed that for the same approximate recovery channel as~\cite{schumacher2001approximate}, the trace distance between the recovered state and the encoded state can be bounded by the mutual purity,
\bea
\bigg|\bigg|\mathcal{R}\left(\rho_{\text{err}}\right)-\rho_{R}\bigg|\bigg|_{1} \leq d_{R}^{5/2}d_{E}^{1/2}\left(\mathcal{F}_{\Psi^{\prime}}(R:E)\right)^{1/4}.
\label{eq:approx_qec2}
\eea

We provide a description of the recovery channel and the sketch of the proof of this bound in Appendix~\ref{appsec:qec_proof}. 
This bound can be computed using just a two-replica computation for local Brownian circuits in $1+1$d with the imaginary TEBD protocol that we have introduced earlier.

\subsection{Numerical results in 1d}

\begin{figure}[htp]
\centering
    \includegraphics[width = \columnwidth]{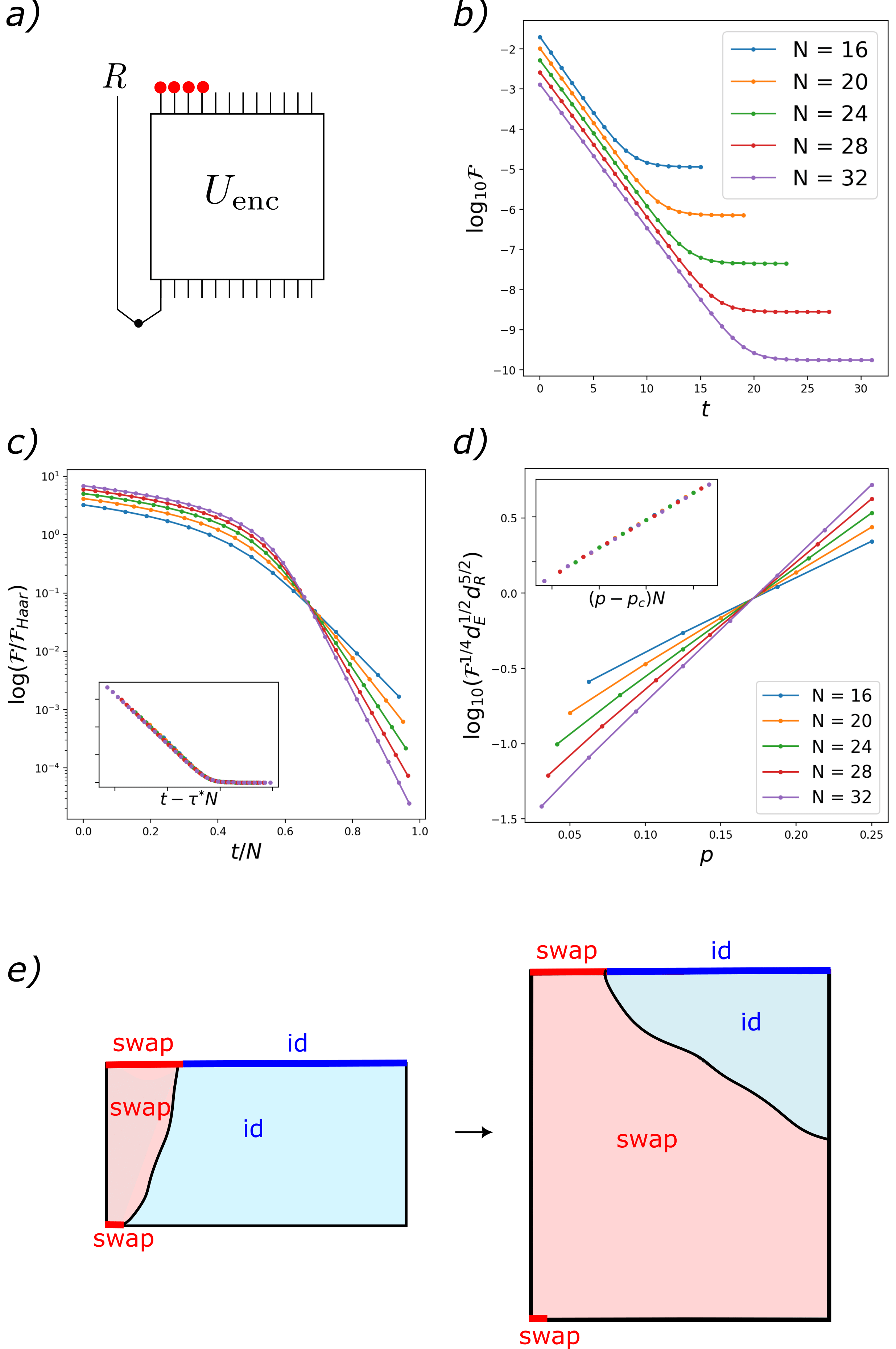}
    \caption{\textbf{2-design transition in the Brownian circuit:} In (a) we encode 1 qubit on the left end, and have erasures (depolarization channel of strength $\lambda = 0.75$) act on $p$ fraction of qubits on the left of the chain. (b) The Mutual purity for the setup in (a) with $p = 0.25$ is plotted as a function of time. (c) $\mathcal{F}$ saturates to $\mathcal{F}_{\text{Haar}} \approx O(2^{-N})$ after $t\sim O(N)$ time. We scale the time linearly with $N$ and find that there is a transition at $t = \tau^{*} N$, with $\tau^{*}\approx 0.77$. The scaling collapse of this transition is shown in the inset. (d) We plot the RHS of the error correction bound of Eq.~\ref{eq:approx_qec2} for different values $p$, with the saturated mutual purity after $t\sim N$ time of encoding by the Brownian circuit. There is a threshold transition $p_{c} \approx 0.17$, and below this fraction, the error is correctable. The scaling collapse of this transition is shown in the inset. (e) The transition in the mutual purity can be related to a first order pinning transition between two domain wall configurations separating domains of two ordered states consistent with the boundary conditions set by the definition of mutual purity. As described in the text, the trace structure in the definition of mutual purity forces the boundary state to certain ordered states. In the initial time boundary, the state is `swap' ordered in the part where the reference qubit is encoded (red), and has free boundary condition elsewhere (no ordered state). The time-like boundaries on the left and right are also free by the open boundary condition. In the final state, the `swap' (red) part refers to the erasure errors, and the `id' (blue) part refers to the partial trace in the definition of mutual purity.}
    \label{fig:qecc_1}
\end{figure}

\begin{figure}[htp]
\centering
    \includegraphics[width = \columnwidth]{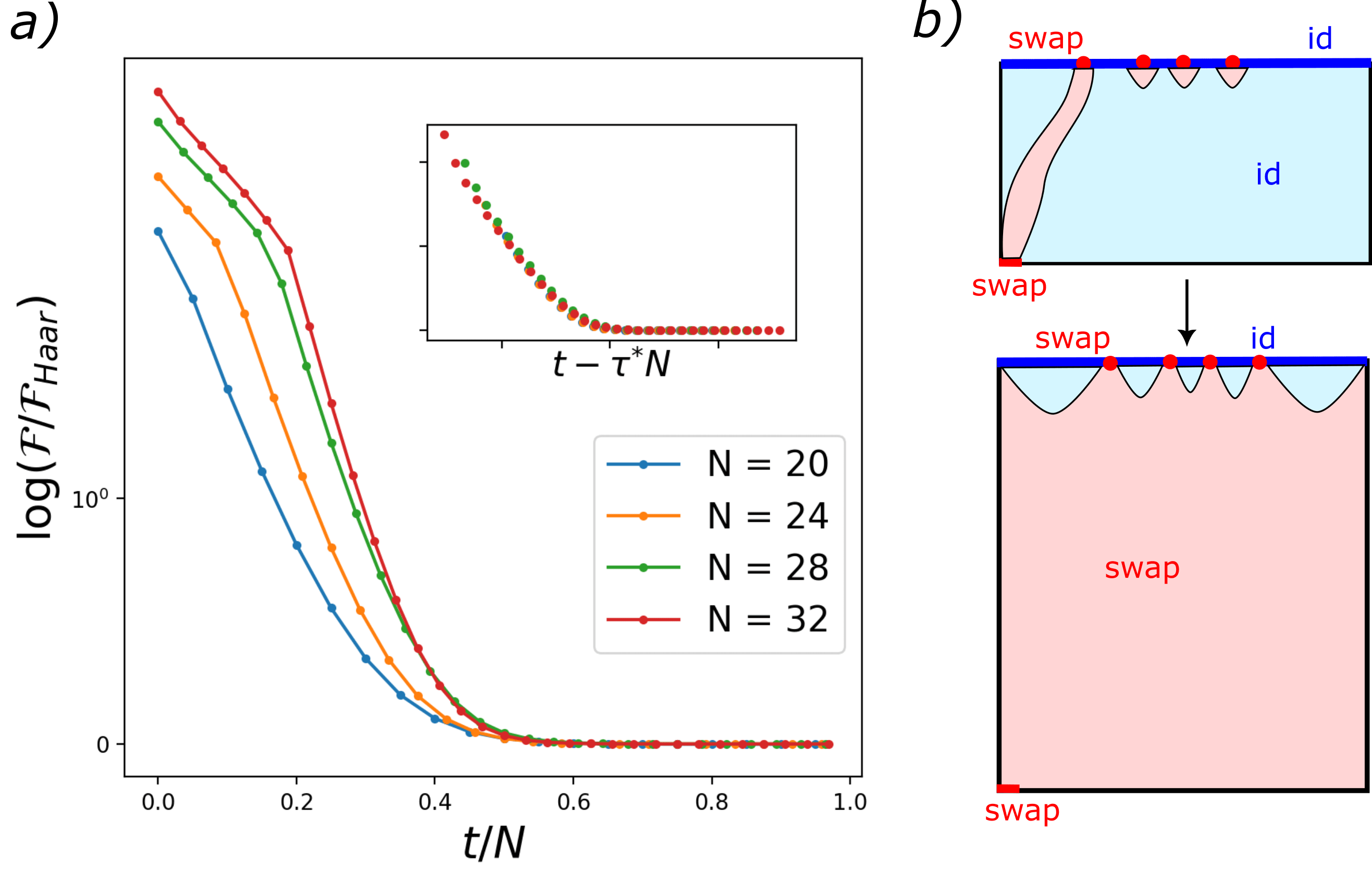}
    \caption{Mutual Purity with erasures of $p = 0.25$ fraction of qubits on random locations. The data presented is averaged over 80 (20) random instances of the erasure locations for $N = 20, 24, 28, (32)$ chains.}
    \label{fig:qecc_3}
\end{figure}

\begin{figure}[htp]
\centering
    \includegraphics[width = \columnwidth]{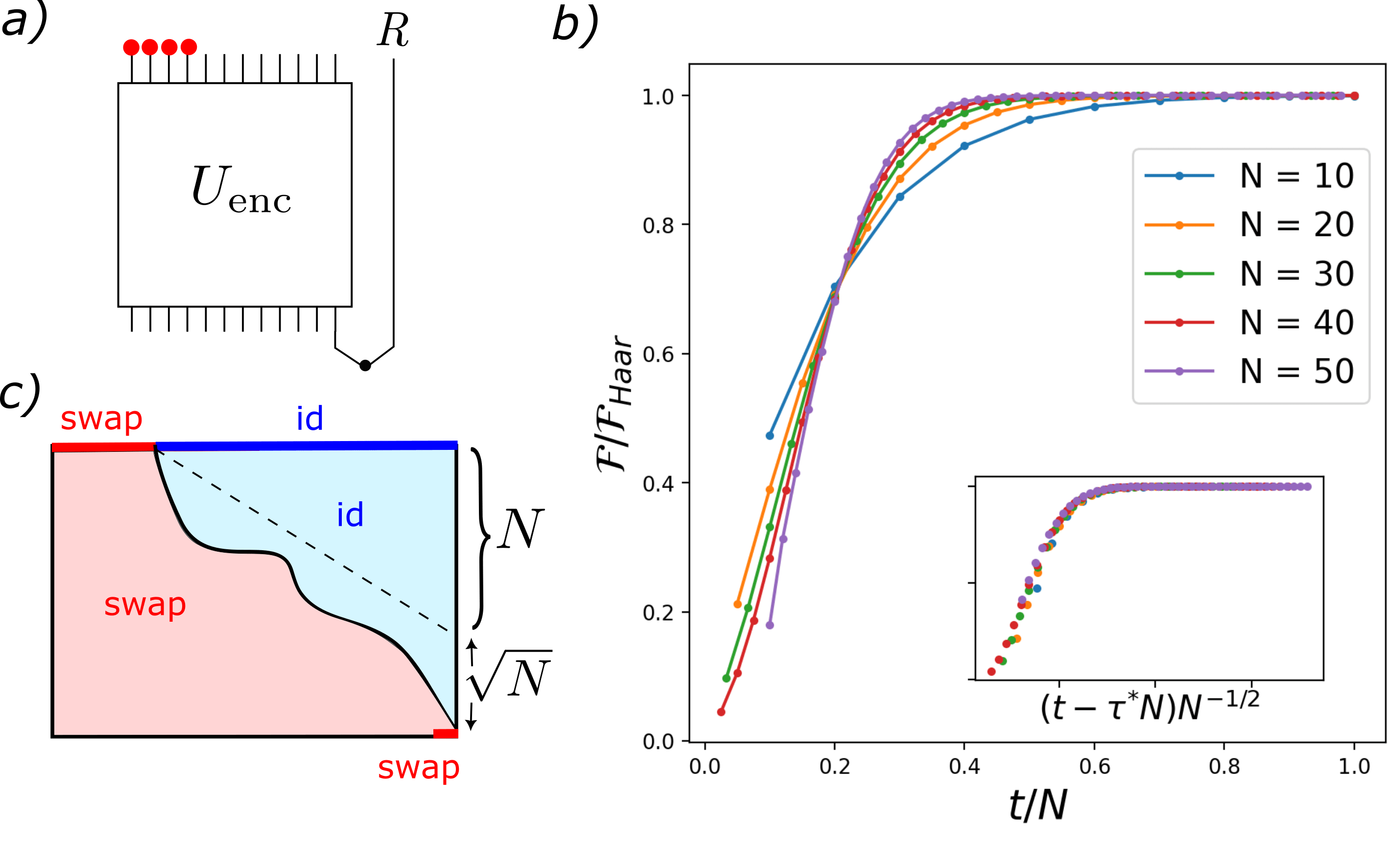}
    \caption{A single qubit reference is encoded on right end of chain, while erasures of $p = 0.25$ fraction of qubits on left of chain (setup shown in (a)). Only a specific domain wall configuration (c) contributes to the mutual purity dynamics as plotted in (b). }
    \label{fig:qecc_4}
\end{figure}

Using the replicated Hilbert space formalism, we can represent the mutual purity for the $1+1$d local Brownian circuit by the following expression,
\bea
\mathcal{F}^{\Psi^{\prime}}_{RE} = \langle\langle \psi_{\text{err}}|\mathbb{U}_{\text{enc}}|\psi_{\text{in}}\rangle \rangle = \langle\langle \psi_{\text{err}}|e^{-t H_{\text{eff}}}|\psi_{\text{in}}\rangle \rangle,
\label{eq:mutual_purity_replica}
\eea
where $\mathbb{U}_{\text{enc}} = \mathbb{E}\left[U_{\text{enc}}\otimes U^{\dagger}_{\text{enc}} \otimes U_{\text{enc}} \otimes U^{\dagger}_{\text{enc}}\right]$, and appropriately defined states $\kett{\psi_{\text{in}}}, \kett{\psi_{\text{err}}}$ in the replicated Hilbert space, given by, 

\begin{align}
    &|\psi_{\text{in}}\rangle \rangle = \left(\kettswap - \frac{1}{2} \kettid\right)^{\otimes A_{1}} \otimes\kett{0000}^{\otimes A_{2}} \\ 
    &|\psi_{\text{err}}\rangle \rangle = 2^{N}\sum_{m,n=0}^{d_{E}-1} E_{m}E_{n}^{*}E_{n}E_{m}^{*}\kettid^{\otimes A}.
\end{align} 
Notice that both $|\psi_{\text{in}}\rangle \rangle$ $|\psi_{\text{err}}\rangle \rangle$ are not normalized.

The operators $E_{m}$ are non-unitary operators implementing the error on the systems $A$,
\bea 
U_{\text{err}}\ket{\psi}_{A}\ket{e_{0}}_{E} = \sum_{m} E^{m}_{A}.
\eea

The derivation is provided in Appendix~\ref{appsec:qec_mp}. The replica order of the initial state $\kett{\psi_{\text{in}}}$ reveals that the state breaks the replica symmetry to `swap' in the region $A_{1}$ (reflecting the encoded qubit), and preserves the replica symmetry in $A_{2}$. As for the final state, $\kett{\psi_{\text{err}}}$, the replica order is `id' in the region where the error doesn't act, and `swap' in the region where error acts.

To diagnose the error correcting properties of the Brownian circuit, we need to take specific noise models. 
In this section, we focus on local depolarization channels acting on a few qubits, say a fraction $p$ of them. 
The depolarization channel of strength $\lambda$ acts on the density matrix as follows,
\bea
\mathcal{N}_{i}(\rho) = (1-\lambda) \rho + \frac{\lambda}{3}\left(\sum_{\alpha = x,y,z}\sigma_{i,\alpha}\rho\sigma_{i,\alpha}\right).
\eea

In Fig.~\ref{fig:qecc_1}a we present the plot of the mutual purity of the $1+1$d Brownian circuit as a function of time, where a single qubit in $R$ is encoded in the system $A$ of size $N$. 
The noise model is chosen to depolarization channel of $\lambda = 0.05$ acting on $p = 0.25$ fraction of qubits. 
It is clear from the plot that the mutual purity initially exponentially decays, until it saturates to the global Haar value which is $O(2^{-N})$. 
The time taken for the saturation scales as $t\propto N$.   
In Appendix~\ref{appsec:qec_haar} we derive the explicit result for mutual purity with globally Haar random encoding $\mathcal{F}_{\text{Haar}} = O(2^{-N})$. 
This numerical result demonstrates that the Brownian circuit approximates a two design in $O(N)$ times, and we show in Fig.~\ref{fig:qecc_1}b that the 2 design transition occurs after time $\tau^{*}N$, where $\tau^{*}\sim 0.77$. The scaling collapse of the transition reveals $\mathcal{F}/\mathcal{F}_{\text{Haar}} \sim f(t-\tau^{*}N)$.

Furthermore, we can study the mutual purity and the RHS of the quantum error correction bound Eq.~\ref{eq:approx_qec2} for different values of $p$. 
In Fig.~\ref{fig:qecc_1}d we plot the saturation value of the RHS of Eq.~\ref{eq:approx_qec2} (after the Brownian circuit has run for $t = N$ steps) for different values of $p$ and system sizes $N$, for a single qubit encoding, and depolarization channel of strength $\lambda = 0.05$. 
We find that the RHS of the error correction bound undergoes a transition at $p^{*}\approx 0.17$, which can be identified as the threshold of this quantum error correction code. 
Note the quantum error correction bound in Eq.~\ref{eq:approx_qec2} guarantees that for $p<p^{*}$ the Brownian circuit generates a quantum error correction code whose errors are correctable using the recovery channel outlined in Appendix~\ref{appsec:qec_proof}.

We don't expect this threshold to be tight, as the error correction bound with the mutual purity is expected to be looser than the bound from mutual information. 
However, the numerical results strongly indicate that the quantum error correction transition with $t$, the depth in the Brownian circuit (the time when the circuit approximates a 2 design) and the threshold transition both correspond to a first-order domain wall pinning transition.

\subsection{Coding transitions}
%\textcolor{red}{Need to understand better.}

As discussed in the previous section, the mutual purity is given by the amplitude, $\mathcal{F}^{\Psi^{\prime}}_{RE} = \langle\langle \psi_{\text{err}}|\mathbb{U}_{\text{enc}}|\psi_{\text{in}}\rangle \rangle $. 
It is convenient to view the space-time layout of the Brownian circuit as a two-dimensional statistical model.
In our setting, this is nothing but the mapping from a $d$ dimensional quantum system to a $d+1$ dimensional classical system.
It is important to note that in the wave function $\left|\psi_\text{in} \right> \rangle$, the encoded $|A_1|$ qubits is mapped to a projection to $\left| \text{swap} \right> \rangle$, i.e., 
\bea \label{eq:proj}
    \langle \left< \text{id} \right| \left(\kettswap - \frac{1}{2} \kettid\right) = 0.
\eea
Whereas the wave function of the rest $|A_2|$ qubits behaves as a free boundary condition, i.e.,  $\langle \left< \text{id} |0000 \right> \rangle = \langle \left< \text{swap} |0000 \right> \rangle = \frac12 $.

On the other hand,  $\left|\psi_\text{err} \right> \rangle$ can effectively change the boundary condition on the top layer. 
In particular, for the single-qubit depolarization channel at site $i$, the wave function becomes a superposition of two spins, 
\bea
    \left| \psi_{\text{err},i} \right> \rangle = (1- \frac43 \lambda)^2 \left| \text{id} \right> \rangle + \frac43 \lambda( 1 - \frac23 \lambda )  \left| \text{swap} \right> \rangle.
\eea
Note that when $\lambda = \frac34$, the wave function is given by $\left| \text{swap} \right> \rangle$ only.
Therefore, the statistical mechanical picture is that in the symmetry breaking phase of the Ising model, the boundary condition caused by the noise channel $\left|\psi_\text{err} \right> \rangle$ will induce different domains in the bulk.
Namely, these are domains denoted by either `id' or `swap' (equivalently, the two Ising values) as shown schematically in Fig.~\ref{fig:qecc_1}(e). 
The mutual purity is only nonzero when the encoded qubit is located in the 'swap' domain.

To understand better the coding transition, we perform a finite scaling analysis of mutual purity as a function of depth and discussion different cases in the following.

\begin{itemize}
    \item {\bf Noisy region overlaps the encoding qubit.} As shown in Fig.~\ref{fig:qecc_1}a, the reference qubit is encoded on the left-most edge, and the noise occurs in a contiguous region that is also on the left edge. 
    In this case, there are many domain wall configurations that can contribute to the mutual purity. 
    To simplify the discussion, we focus on two different domain walls: one ends on the bottom layer, and the other ends on the right edge. 
    A schematic plot of these two domain walls is shown in Fig.~\ref{fig:qecc_1}e.
    It is clear that their contributions are 
    \bea
        \mathcal F &\sim& e^{- \Delta t} + e^{-\Delta'(1-p) L} \\
        &= & e^{-\Delta'(1-p) L} (1 + e^{- \Delta t +\Delta'(1-p) L}),
    \eea
    where $\Delta$ and $\Delta'$ denote the tension of the two kinds of domain walls respectively.
    $L$ is the length of the chain. 
    In the short-time region, the first kind of domain wall dominates, while in the long-time region, the second kind of domain wall dominates, and it becomes time independent.
    There is an exchange of dominating domain configurations, as demonstrated in Fig.~\ref{fig:qecc_1}e. 
    The transition time is roughly $\frac{\Delta'(1-p)}{\Delta} L \propto N$. 
    This explains the behavior in Fig.~\ref{fig:qecc_1}b and c. 
    Replacing the contribution from the second kind of domain wall by $\mathcal F_\text{Haar}$, we can obtain $\mathcal F/ \mathcal F_\text{err} = 1 + e^{- (\Delta t + \log \mathcal F_\text{err}) } $, which is consistent with the data collapse performed in the Fig.~\ref{fig:qecc_1}c.
    
    \item {\bf Random noisy region.} In this case, the noise occurs in random positions, as shown in Fig.~\ref{fig:qecc_3}. 
    The picture of exchange of two kinds of domain wall configuration is still correct.
    The inset of Fig.~\ref{fig:qecc_3} shows consistent data collapse. %This can also be identified with 

    \item {\bf Noisy region does not overlap the encoding qubit.} 
    The encoding qubit and the noisy region is shown in Fig.~\ref{fig:qecc_4}.
    In the calculation, we set $\lambda = 3/4$. 
    The boundary condition creates a domain wall at the boundary between the noisy qubits and the noiseless qubits. 
    Due to the causality, the back propagation of the domain wall is constrained in an emergent light cone. 
    Thus, the mutual purity is zero (up to an exponentially small number of $N$) when the encoding coding qubit is still outside the back propagating light cone of the domain wall. 
    Moreover, unlike in the previous case where the first kind of domain wall that ends on the bottom can lead to a finite mutual parity, here, only the second kind of domain wall that ends on the right boundary can have a significant contribution to the mutual purity. 
    This is only possible when the back propagating light one hits the right boundary. 
    Therefore, this indicates a dynamical transition at a timescale that is proportional to the system size.
    In Fig.~\ref{fig:qecc_4}, the crossing of mutual parity in different sizes indicates such a transition.
    
    We also performed the data collapse in the inset of Fig.~\ref{fig:qecc_4}. 
    Different from the previous two cases, the scaling is given by $(t - \tau^\ast N)/\sqrt{N}$.
    To understand this, note that the `id' domain back propagates to the light cone as shown in. 
    In the symmetry breaking phase, the domain wall can fluctuate away from the light cone~\footnote{Note that in the symmetry breaking phase, there are still two phases for domain walls, the pinning and depinning phases, separating by a pinning transition. For our case, since the coupling at the top layer is the same as the coupling in the bulk, the depinning transition is the same as the symmetry breaking transition. This means the domain wall is depinned and can fluctuate.}.
    The average position is $\sqrt{N}$ away from the domain wall~\cite{abraham1980solvable}.
    Furthermore, the fluctuation of the domain wall is captured by a universal function of $\alpha = \delta L/ \sqrt{N}$~\cite{abraham1980solvable}, where $\delta L$ denotes the distance away from the light cone.
    More concretely, the magnetization profile is a function of $\alpha$ at the distance that is of the order $\sqrt N$ (outside this distance, the magnetization is given either by one of the two spin polarization).
    We expect that this function also captures the mutual purity because mutual purity in this case prob the `swap' spin at the right boundary. 
    Now, since the light cone reaches the right boundary at a time of order $N$, the mutual purity is then a universal function of $(t- \frac{N}v)/\sqrt{N}$, where $v$ is the light speed.
    This explains the data collapse.
    
\end{itemize}

In summary, depending on whether the noise  occurs in the encoding qubit, we discover distinct coding transitions.
If the noisy region covers the encoding qubit, there are two kinds of domain wall configurations contributing to the mutual purity.
They are schematically shown in Fig.~\ref{fig:qecc_1}e.
The exchange of dominance between the two kinds of domain walls underlies the physics of the coding transition in this situation. 
On the other hand, if the noisy region does not cover the encoding qubit, the mutual purity is only nonzero when the noise back propagates to the encoding qubit.
In this case, the transition is induced by the fluctuating domain walls 
and captured by a different scaling, $N^{-1/2}$, as shown in the data collapse.

\section{Concluding remarks} \label{sec:discussions}

In this paper, we have used the effective replica Hamiltonian mapping for local Brownian circuits to probe timescales of complexity growth in random quantum circuits, namely anti-concentration and approximate unitary-design generation. 
The effective replica model serves two purposes: we can perform large-scale numerics to simulate several quantum informational quantities for long times, using tensor network tools. 
This makes local Brownian circuits as efficient numerical tools to study unitary quantum many-body dynamics. 
Secondly, it transforms the question of time-scales in the real-time dynamics into questions of energy-scales in a corresponding thermodynamic problem, which allows us to make analytical progress. 

We have shown that local Brownian circuits in $1+1$d anticoncentrate in $\log N$ time, consistent with earlier results in local Haar random circuits~\cite{Dalzell_2022}. 
Furthermore, we have analyzed the success condition of an approximate classical algorithm~\cite{Napp_2022} to sample from the output distribution of Brownian circuits, and have identified that there is a sharp transition in the computational hardness of simulation at the same timescale. 
The anticoncentration transition arises from the transition in dominance of different low-energy states of the effective Hamiltonian in the collision probability. 
In particular, the collision probability (a probe of anticoncentration) gets contribution from eigenstates of the effective Hamiltonian with domain walls and the timescale where this becomes relevant can be related to the logarithm of the number of such domain wall states (which in $1d$ is $\sim N$). 

In the presence of noise, we showed that there is a noise-induced phase transition in linear cross entropy benchmark ($\chi_{\text{XEB}}$), as has been recently demonstrated for related noisy random circuit models in~\cite{morvan2023phase, ware2023sharp}. 
This can be seen as a consequence of explicit replica symmetry breaking in the effective Hamiltonian model in the presence of noise acting on a single replica of the system. 
By relating the $\chi_{\text{XEB}}$ to specific transition amplitudes in the corresponding replica model, we identify the noise-induced transition in the cross entropy benchmarking as the transition in the dominance of certain domain wall states in the presence of explicit bulk symmetry breaking field. 
The critical properties of the transition can be related to those of external field-induced first-order transitions in the classical Ising model in $2d$~\cite{binder1984finite}.

Finally, we probed the generation of approximate unitary design by Brownian circuits. 
By directly probing the quantum error-correcting properties of the Brownian circuit, namely a 2-replica quantity called Mutual Purity~\cite{balasubramanian2023quantum}, we find that the $1+1$d Brownian circuits become good quantum error correcting codes in $O(N)$ time. 
This transition can be identified as first-order transitions between certain space-time domain wall configurations, which are related to first-order boundary-driven pinning transitions in classical Ising models.

There can be several avenues of future research based on this work. 
Here we have demonstrated $1+1$d Brownian circuits as a useful numerically accessible tool for studying the dynamics of quantum information. 
A natural question is whether the same numerical feasibility extends to higher dimensions. 
%An intriguing question is whether the existence of a thermal transition in $2+1d$ and higher circuits correspond to observable impact on any quantum-informational quantity. 
Here, we speculate the dynamics and transitions in informational quantities in higher dimensional Brownian circuits $d>1$ (here $d$ is the spatial dimension of the Brownian circuit, $N$ is total number of qubits, we also use the volume $L^d \sim N$ with $L$ denotes the length scale):
\begin{itemize}
    \item {\bf Collision probability}.
    It is still true that in higher dimensions, the collision probability at long enough time is dominated by two grounds states and elementary excitations. 
    Distinct from the situation in 1d, now the lowest excitation is given by local spin flips with an energy that is independent of system sizes.
    Nevertheless, the entropy of such a local excitation is proportional to the system size $N$.
    Therefore, we expect that the Brownian circuit anti-concentrates on a $\log N$ timescale, similar to that in 1d~\cite{Dalzell_2022}.
    
    \item{\bf Computational transition in patching algorithm}. 
    The patching algorithm is closely related to the symmetry breaking of the underlying 2-replica spin model~\cite{Napp_2022}.
    In higher dimensions, $d>1$, the discrete symmetry can be broken in a finite depth. 
    This contrasts with the 1d case, where even the discrete symmetry can only be broken in a $\log$ depth.
    This means that the patching algorithm will fail when the depth of the Brownian circuit exceeds a critical depth that is independent of system sizes.

    \item {\bf Noisy cross-entropy benchmarking}. 
    A noise $\epsilon$ behaves as an external field and will lift the degeneracy between $\left| \text{id} \right>\rangle $ and $\left| \text{swap} \right>\rangle $, i.e., $\left| \text{swap} \right>\rangle $ will be suppressed by a factor of $e^{- N \epsilon t}$.
    On the other hand, as discussed in the collision probability, the elementary excitation is given by local spin flips. 
    With an external field, there is an additional cost, adding up to a factor $\sim e^{- z \Delta t - \epsilon t} $, where $z$ is the coordination number.
    Therefore, we expect that when the noise rate scales as $1/N = 1/L^d$, there will be a first-order phase transition with critical exponent given by $1/\nu = d+1$.

    \item {\bf Coding transition}.
    The dynamical transition for the Brownian circuit to achieve an approximate unitary 2-design is given by the transition of dominance between two kinds of domain walls.
    It should be the same in higher dimensions.
    Therefore, the transition occurs on a timescale $\sim L = N^{1/d}$~\cite{Harrow2023approximate}.
    Next, consider the different regions of noise and the encoding qubit. 
    We expect the mutual purity transition is similarly given by transitions of domain walls when the noisy region overlaps the encoding qubit.
    On the other hand, when the noisy region does not overlap the encoding qubit, we expect the fluctuation of domain wall dictates the coding transition.
\end{itemize}

Even in $1+1$d, this work paves the way towards exploration of quantum information dynamics in symmetric Brownian circuits, by studying directly the spectrum of the effective Hamiltonian in the presence of other circuit symmetries. 
Another direction of interest is incorporating the effects of mid-circuit measurements in the entanglement dynamics in Brownian circuit~\cite{jian2021measurement,jian2021phase}. 
We will present these results in a future work.

In this work, we have focused on only 2-replica quantities, such as collision probabilities and mutual purities. 
In principle, any integer $k$ replica quantities can be represented in the effective Hamiltonian picture, with $q^{k}$ local Hilbert space dimension ($q$ being the dimension of the original degrees of freedom), which makes numerical methods intractable at large sizes for large $k$. 
An outstanding question is to develop controlled numerical methods or analytical techniques to take the $k\to 1$ replica limit. 

\section*{Acknowledgements}

We thank Timothy Hsieh, Tsung-Cheng Lu, Utkarsh Agrawal, and Xuan Zou for useful discussions, and Tsung-Cheng Lu for detailed comments. We have used the TeNPy package for the tensor network simulations~\cite{tenpy}.
The numerical simulations were performed using the Symmetry HPC system at Perimeter Institute (PI). Research at PI is supported in part by the Government of Canada through the Department of Innovation, Science and Economic Development and by the Province of Ontario through the Ministry of Colleges and Universities.
S.-K.J is supported by a startup fund at Tulane University.

\bibliography{references.bib}

\begin{thebibliography}{52}%
\makeatletter
\providecommand \@ifxundefined [1]{%
 \@ifx{#1\undefined}
}%
\providecommand \@ifnum [1]{%
 \ifnum #1\expandafter \@firstoftwo
 \else \expandafter \@secondoftwo
 \fi
}%
\providecommand \@ifx [1]{%
 \ifx #1\expandafter \@firstoftwo
 \else \expandafter \@secondoftwo
 \fi
}%
\providecommand \natexlab [1]{#1}%
\providecommand \enquote  [1]{``#1''}%
\providecommand \bibnamefont  [1]{#1}%
\providecommand \bibfnamefont [1]{#1}%
\providecommand \citenamefont [1]{#1}%
\providecommand \href@noop [0]{\@secondoftwo}%
\providecommand \href [0]{\begingroup \@sanitize@url \@href}%
\providecommand \@href[1]{\@@startlink{#1}\@@href}%
\providecommand \@@href[1]{\endgroup#1\@@endlink}%
\providecommand \@sanitize@url [0]{\catcode `\\12\catcode `\$12\catcode
  `\&12\catcode `\#12\catcode `\^12\catcode `\_12\catcode `\%12\relax}%
\providecommand \@@startlink[1]{}%
\providecommand \@@endlink[0]{}%
\providecommand \url  [0]{\begingroup\@sanitize@url \@url }%
\providecommand \@url [1]{\endgroup\@href {#1}{\urlprefix }}%
\providecommand \urlprefix  [0]{URL }%
\providecommand \Eprint [0]{\href }%
\providecommand \doibase [0]{https://doi.org/}%
\providecommand \selectlanguage [0]{\@gobble}%
\providecommand \bibinfo  [0]{\@secondoftwo}%
\providecommand \bibfield  [0]{\@secondoftwo}%
\providecommand \translation [1]{[#1]}%
\providecommand \BibitemOpen [0]{}%
\providecommand \bibitemStop [0]{}%
\providecommand \bibitemNoStop [0]{.\EOS\space}%
\providecommand \EOS [0]{\spacefactor3000\relax}%
\providecommand \BibitemShut  [1]{\csname bibitem#1\endcsname}%
\let\auto@bib@innerbib\@empty
%</preamble>
\bibitem [{\citenamefont {Nahum}\ \emph {et~al.}(2018)\citenamefont {Nahum},
  \citenamefont {Vijay},\ and\ \citenamefont {Haah}}]{nahum2018operator}%
  \BibitemOpen
  \bibfield  {author} {\bibinfo {author} {\bibfnamefont {A.}~\bibnamefont
  {Nahum}}, \bibinfo {author} {\bibfnamefont {S.}~\bibnamefont {Vijay}},\ and\
  \bibinfo {author} {\bibfnamefont {J.}~\bibnamefont {Haah}},\ }\bibfield
  {journal} {\bibinfo  {journal} {Physical Review X}\ }\textbf {\bibinfo
  {volume} {8}},\ \href {https://doi.org/10.1103/physrevx.8.021014}
  {10.1103/physrevx.8.021014} (\bibinfo {year} {2018})\BibitemShut {NoStop}%
\bibitem [{\citenamefont {von Keyserlingk}\ \emph {et~al.}(2018)\citenamefont
  {von Keyserlingk}, \citenamefont {Rakovszky}, \citenamefont {Pollmann},\ and\
  \citenamefont {Sondhi}}]{von2018operator}%
  \BibitemOpen
  \bibfield  {author} {\bibinfo {author} {\bibfnamefont {C.}~\bibnamefont {von
  Keyserlingk}}, \bibinfo {author} {\bibfnamefont {T.}~\bibnamefont
  {Rakovszky}}, \bibinfo {author} {\bibfnamefont {F.}~\bibnamefont
  {Pollmann}},\ and\ \bibinfo {author} {\bibfnamefont {S.}~\bibnamefont
  {Sondhi}},\ }\bibfield  {journal} {\bibinfo  {journal} {Physical Review X}\
  }\textbf {\bibinfo {volume} {8}},\ \href
  {https://doi.org/10.1103/physrevx.8.021013} {10.1103/physrevx.8.021013}
  (\bibinfo {year} {2018})\BibitemShut {NoStop}%
\bibitem [{\citenamefont {Zhou}\ and\ \citenamefont
  {Nahum}(2019)}]{zhou2019emergent}%
  \BibitemOpen
  \bibfield  {author} {\bibinfo {author} {\bibfnamefont {T.}~\bibnamefont
  {Zhou}}\ and\ \bibinfo {author} {\bibfnamefont {A.}~\bibnamefont {Nahum}},\
  }\bibfield  {journal} {\bibinfo  {journal} {Physical Review B}\ }\textbf
  {\bibinfo {volume} {99}},\ \href {https://doi.org/10.1103/physrevb.99.174205}
  {10.1103/physrevb.99.174205} (\bibinfo {year} {2019})\BibitemShut {NoStop}%
\bibitem [{\citenamefont {Khemani}\ \emph {et~al.}(2018)\citenamefont
  {Khemani}, \citenamefont {Vishwanath},\ and\ \citenamefont
  {Huse}}]{khemani2018operator}%
  \BibitemOpen
  \bibfield  {author} {\bibinfo {author} {\bibfnamefont {V.}~\bibnamefont
  {Khemani}}, \bibinfo {author} {\bibfnamefont {A.}~\bibnamefont
  {Vishwanath}},\ and\ \bibinfo {author} {\bibfnamefont {D.~A.}\ \bibnamefont
  {Huse}},\ }\href@noop {} {\bibfield  {journal} {\bibinfo  {journal} {Physical
  Review X}\ }\textbf {\bibinfo {volume} {8}},\ \bibinfo {pages} {031057}
  (\bibinfo {year} {2018})}\BibitemShut {NoStop}%
\bibitem [{\citenamefont {Rakovszky}\ \emph {et~al.}(2018)\citenamefont
  {Rakovszky}, \citenamefont {Pollmann},\ and\ \citenamefont
  {Von~Keyserlingk}}]{rakovszky2018diffusive}%
  \BibitemOpen
  \bibfield  {author} {\bibinfo {author} {\bibfnamefont {T.}~\bibnamefont
  {Rakovszky}}, \bibinfo {author} {\bibfnamefont {F.}~\bibnamefont
  {Pollmann}},\ and\ \bibinfo {author} {\bibfnamefont {C.}~\bibnamefont
  {Von~Keyserlingk}},\ }\href@noop {} {\bibfield  {journal} {\bibinfo
  {journal} {Physical Review X}\ }\textbf {\bibinfo {volume} {8}},\ \bibinfo
  {pages} {031058} (\bibinfo {year} {2018})}\BibitemShut {NoStop}%
\bibitem [{\citenamefont {Hayden}\ \emph {et~al.}(2016)\citenamefont {Hayden},
  \citenamefont {Nezami}, \citenamefont {Qi}, \citenamefont {Thomas},
  \citenamefont {Walter},\ and\ \citenamefont {Yang}}]{hayden2016holographic}%
  \BibitemOpen
  \bibfield  {author} {\bibinfo {author} {\bibfnamefont {P.}~\bibnamefont
  {Hayden}}, \bibinfo {author} {\bibfnamefont {S.}~\bibnamefont {Nezami}},
  \bibinfo {author} {\bibfnamefont {X.-L.}\ \bibnamefont {Qi}}, \bibinfo
  {author} {\bibfnamefont {N.}~\bibnamefont {Thomas}}, \bibinfo {author}
  {\bibfnamefont {M.}~\bibnamefont {Walter}},\ and\ \bibinfo {author}
  {\bibfnamefont {Z.}~\bibnamefont {Yang}},\ }\href@noop {} {\bibfield
  {journal} {\bibinfo  {journal} {Journal of High Energy Physics}\ }\textbf
  {\bibinfo {volume} {2016}},\ \bibinfo {pages} {1} (\bibinfo {year}
  {2016})}\BibitemShut {NoStop}%
\bibitem [{\citenamefont {Arute}\ \emph {et~al.}(2019)\citenamefont {Arute},
  \citenamefont {Arya}, \citenamefont {Babbush}, \citenamefont {Bacon},
  \citenamefont {Bardin}, \citenamefont {Barends}, \citenamefont {Biswas},
  \citenamefont {Boixo}, \citenamefont {Brandao}, \citenamefont {Buell} \emph
  {et~al.}}]{arute2019quantum}%
  \BibitemOpen
  \bibfield  {author} {\bibinfo {author} {\bibfnamefont {F.}~\bibnamefont
  {Arute}}, \bibinfo {author} {\bibfnamefont {K.}~\bibnamefont {Arya}},
  \bibinfo {author} {\bibfnamefont {R.}~\bibnamefont {Babbush}}, \bibinfo
  {author} {\bibfnamefont {D.}~\bibnamefont {Bacon}}, \bibinfo {author}
  {\bibfnamefont {J.~C.}\ \bibnamefont {Bardin}}, \bibinfo {author}
  {\bibfnamefont {R.}~\bibnamefont {Barends}}, \bibinfo {author} {\bibfnamefont
  {R.}~\bibnamefont {Biswas}}, \bibinfo {author} {\bibfnamefont
  {S.}~\bibnamefont {Boixo}}, \bibinfo {author} {\bibfnamefont {F.~G.}\
  \bibnamefont {Brandao}}, \bibinfo {author} {\bibfnamefont {D.~A.}\
  \bibnamefont {Buell}}, \emph {et~al.},\ }\href@noop {} {\bibfield  {journal}
  {\bibinfo  {journal} {Nature}\ }\textbf {\bibinfo {volume} {574}},\ \bibinfo
  {pages} {505} (\bibinfo {year} {2019})}\BibitemShut {NoStop}%
\bibitem [{\citenamefont {Morvan}\ \emph {et~al.}(2023)\citenamefont {Morvan},
  \citenamefont {Villalonga}, \citenamefont {Mi}, \citenamefont {Mandr{\`a}},
  \citenamefont {Bengtsson}, \citenamefont {Klimov}, \citenamefont {Chen},
  \citenamefont {Hong}, \citenamefont {Erickson}, \citenamefont {Drozdov} \emph
  {et~al.}}]{morvan2023phase}%
  \BibitemOpen
  \bibfield  {author} {\bibinfo {author} {\bibfnamefont {A.}~\bibnamefont
  {Morvan}}, \bibinfo {author} {\bibfnamefont {B.}~\bibnamefont {Villalonga}},
  \bibinfo {author} {\bibfnamefont {X.}~\bibnamefont {Mi}}, \bibinfo {author}
  {\bibfnamefont {S.}~\bibnamefont {Mandr{\`a}}}, \bibinfo {author}
  {\bibfnamefont {A.}~\bibnamefont {Bengtsson}}, \bibinfo {author}
  {\bibfnamefont {P.}~\bibnamefont {Klimov}}, \bibinfo {author} {\bibfnamefont
  {Z.}~\bibnamefont {Chen}}, \bibinfo {author} {\bibfnamefont {S.}~\bibnamefont
  {Hong}}, \bibinfo {author} {\bibfnamefont {C.}~\bibnamefont {Erickson}},
  \bibinfo {author} {\bibfnamefont {I.}~\bibnamefont {Drozdov}}, \emph
  {et~al.},\ }\href@noop {} {\bibfield  {journal} {\bibinfo  {journal} {arXiv
  preprint arXiv:2304.11119}\ } (\bibinfo {year} {2023})}\BibitemShut {NoStop}%
\bibitem [{\citenamefont {Wu}\ \emph {et~al.}(2021)\citenamefont {Wu},
  \citenamefont {Bao}, \citenamefont {Cao}, \citenamefont {Chen}, \citenamefont
  {Chen}, \citenamefont {Chen}, \citenamefont {Chung}, \citenamefont {Deng},
  \citenamefont {Du}, \citenamefont {Fan} \emph {et~al.}}]{wu2021strong}%
  \BibitemOpen
  \bibfield  {author} {\bibinfo {author} {\bibfnamefont {Y.}~\bibnamefont
  {Wu}}, \bibinfo {author} {\bibfnamefont {W.-S.}\ \bibnamefont {Bao}},
  \bibinfo {author} {\bibfnamefont {S.}~\bibnamefont {Cao}}, \bibinfo {author}
  {\bibfnamefont {F.}~\bibnamefont {Chen}}, \bibinfo {author} {\bibfnamefont
  {M.-C.}\ \bibnamefont {Chen}}, \bibinfo {author} {\bibfnamefont
  {X.}~\bibnamefont {Chen}}, \bibinfo {author} {\bibfnamefont {T.-H.}\
  \bibnamefont {Chung}}, \bibinfo {author} {\bibfnamefont {H.}~\bibnamefont
  {Deng}}, \bibinfo {author} {\bibfnamefont {Y.}~\bibnamefont {Du}}, \bibinfo
  {author} {\bibfnamefont {D.}~\bibnamefont {Fan}}, \emph {et~al.},\
  }\href@noop {} {\bibfield  {journal} {\bibinfo  {journal} {Physical review
  letters}\ }\textbf {\bibinfo {volume} {127}},\ \bibinfo {pages} {180501}
  (\bibinfo {year} {2021})}\BibitemShut {NoStop}%
\bibitem [{\citenamefont {Zhu}\ \emph {et~al.}(2022)\citenamefont {Zhu},
  \citenamefont {Cao}, \citenamefont {Chen}, \citenamefont {Chen},
  \citenamefont {Chen}, \citenamefont {Chung}, \citenamefont {Deng},
  \citenamefont {Du}, \citenamefont {Fan}, \citenamefont {Gong} \emph
  {et~al.}}]{zhu2022quantum}%
  \BibitemOpen
  \bibfield  {author} {\bibinfo {author} {\bibfnamefont {Q.}~\bibnamefont
  {Zhu}}, \bibinfo {author} {\bibfnamefont {S.}~\bibnamefont {Cao}}, \bibinfo
  {author} {\bibfnamefont {F.}~\bibnamefont {Chen}}, \bibinfo {author}
  {\bibfnamefont {M.-C.}\ \bibnamefont {Chen}}, \bibinfo {author}
  {\bibfnamefont {X.}~\bibnamefont {Chen}}, \bibinfo {author} {\bibfnamefont
  {T.-H.}\ \bibnamefont {Chung}}, \bibinfo {author} {\bibfnamefont
  {H.}~\bibnamefont {Deng}}, \bibinfo {author} {\bibfnamefont {Y.}~\bibnamefont
  {Du}}, \bibinfo {author} {\bibfnamefont {D.}~\bibnamefont {Fan}}, \bibinfo
  {author} {\bibfnamefont {M.}~\bibnamefont {Gong}}, \emph {et~al.},\
  }\href@noop {} {\bibfield  {journal} {\bibinfo  {journal} {Science bulletin}\
  }\textbf {\bibinfo {volume} {67}},\ \bibinfo {pages} {240} (\bibinfo {year}
  {2022})}\BibitemShut {NoStop}%
\bibitem [{\citenamefont {Brown}\ and\ \citenamefont
  {Fawzi}(2013)}]{Brown_2013}%
  \BibitemOpen
  \bibfield  {author} {\bibinfo {author} {\bibfnamefont {W.}~\bibnamefont
  {Brown}}\ and\ \bibinfo {author} {\bibfnamefont {O.}~\bibnamefont {Fawzi}},\
  }in\ \href {https://doi.org/10.1109/isit.2013.6620245} {\emph {\bibinfo
  {booktitle} {2013 {IEEE} International Symposium on Information Theory}}}\
  (\bibinfo  {publisher} {{IEEE}},\ \bibinfo {year} {2013})\BibitemShut
  {NoStop}%
\bibitem [{\citenamefont {Brown}\ and\ \citenamefont
  {Fawzi}(2015)}]{Brown_2015}%
  \BibitemOpen
  \bibfield  {author} {\bibinfo {author} {\bibfnamefont {W.}~\bibnamefont
  {Brown}}\ and\ \bibinfo {author} {\bibfnamefont {O.}~\bibnamefont {Fawzi}},\
  }\href {https://doi.org/10.1007/s00220-015-2470-1} {\bibfield  {journal}
  {\bibinfo  {journal} {Communications in Mathematical Physics}\ }\textbf
  {\bibinfo {volume} {340}},\ \bibinfo {pages} {867} (\bibinfo {year}
  {2015})}\BibitemShut {NoStop}%
\bibitem [{\citenamefont {Gullans}\ \emph {et~al.}(2021)\citenamefont
  {Gullans}, \citenamefont {Krastanov}, \citenamefont {Huse}, \citenamefont
  {Jiang},\ and\ \citenamefont {Flammia}}]{Gullans_QuantumCoding_2021}%
  \BibitemOpen
  \bibfield  {author} {\bibinfo {author} {\bibfnamefont {M.~J.}\ \bibnamefont
  {Gullans}}, \bibinfo {author} {\bibfnamefont {S.}~\bibnamefont {Krastanov}},
  \bibinfo {author} {\bibfnamefont {D.~A.}\ \bibnamefont {Huse}}, \bibinfo
  {author} {\bibfnamefont {L.}~\bibnamefont {Jiang}},\ and\ \bibinfo {author}
  {\bibfnamefont {S.~T.}\ \bibnamefont {Flammia}},\ }\href
  {https://doi.org/10.1103/PhysRevX.11.031066} {\bibfield  {journal} {\bibinfo
  {journal} {Phys. Rev. X}\ }\textbf {\bibinfo {volume} {11}},\ \bibinfo
  {pages} {031066} (\bibinfo {year} {2021})}\BibitemShut {NoStop}%
\bibitem [{\citenamefont {Boixo}\ \emph {et~al.}(2018)\citenamefont {Boixo},
  \citenamefont {Isakov}, \citenamefont {Smelyanskiy}, \citenamefont {Babbush},
  \citenamefont {Ding}, \citenamefont {Jiang}, \citenamefont {Bremner},
  \citenamefont {Martinis},\ and\ \citenamefont {Neven}}]{Boixo_2018}%
  \BibitemOpen
  \bibfield  {author} {\bibinfo {author} {\bibfnamefont {S.}~\bibnamefont
  {Boixo}}, \bibinfo {author} {\bibfnamefont {S.~V.}\ \bibnamefont {Isakov}},
  \bibinfo {author} {\bibfnamefont {V.~N.}\ \bibnamefont {Smelyanskiy}},
  \bibinfo {author} {\bibfnamefont {R.}~\bibnamefont {Babbush}}, \bibinfo
  {author} {\bibfnamefont {N.}~\bibnamefont {Ding}}, \bibinfo {author}
  {\bibfnamefont {Z.}~\bibnamefont {Jiang}}, \bibinfo {author} {\bibfnamefont
  {M.~J.}\ \bibnamefont {Bremner}}, \bibinfo {author} {\bibfnamefont {J.~M.}\
  \bibnamefont {Martinis}},\ and\ \bibinfo {author} {\bibfnamefont
  {H.}~\bibnamefont {Neven}},\ }\href
  {https://doi.org/10.1038/s41567-018-0124-x} {\bibfield  {journal} {\bibinfo
  {journal} {Nature Physics}\ }\textbf {\bibinfo {volume} {14}},\ \bibinfo
  {pages} {595} (\bibinfo {year} {2018})}\BibitemShut {NoStop}%
\bibitem [{\citenamefont {Hangleiter}\ and\ \citenamefont
  {Eisert}(2023)}]{hangleiter2023computational}%
  \BibitemOpen
  \bibfield  {author} {\bibinfo {author} {\bibfnamefont {D.}~\bibnamefont
  {Hangleiter}}\ and\ \bibinfo {author} {\bibfnamefont {J.}~\bibnamefont
  {Eisert}},\ }\href@noop {} {\bibinfo {title} {Computational advantage of
  quantum random sampling}} (\bibinfo {year} {2023}),\ \Eprint
  {https://arxiv.org/abs/2206.04079} {arXiv:2206.04079 [quant-ph]} \BibitemShut
  {NoStop}%
\bibitem [{\citenamefont {Dalzell}\ \emph {et~al.}(2022)\citenamefont
  {Dalzell}, \citenamefont {Hunter-Jones},\ and\ \citenamefont {Brand{\~{a}
  }o}}]{Dalzell_2022}%
  \BibitemOpen
  \bibfield  {author} {\bibinfo {author} {\bibfnamefont {A.~M.}\ \bibnamefont
  {Dalzell}}, \bibinfo {author} {\bibfnamefont {N.}~\bibnamefont
  {Hunter-Jones}},\ and\ \bibinfo {author} {\bibfnamefont {F.~G. S.~L.}\
  \bibnamefont {Brand{\~{a} }o}},\ }\bibfield  {journal} {\bibinfo  {journal}
  {{PRX} Quantum}\ }\textbf {\bibinfo {volume} {3}},\ \href
  {https://doi.org/10.1103/prxquantum.3.010333} {10.1103/prxquantum.3.010333}
  (\bibinfo {year} {2022})\BibitemShut {NoStop}%
\bibitem [{\citenamefont {Brand{\~{a}}o}\ \emph {et~al.}(2016)\citenamefont
  {Brand{\~{a}}o}, \citenamefont {Harrow},\ and\ \citenamefont
  {Horodecki}}]{Brandao_2016_local}%
  \BibitemOpen
  \bibfield  {author} {\bibinfo {author} {\bibfnamefont {F.~G. S.~L.}\
  \bibnamefont {Brand{\~{a}}o}}, \bibinfo {author} {\bibfnamefont {A.~W.}\
  \bibnamefont {Harrow}},\ and\ \bibinfo {author} {\bibfnamefont
  {M.}~\bibnamefont {Horodecki}},\ }\href
  {https://doi.org/10.1007/s00220-016-2706-8} {\bibfield  {journal} {\bibinfo
  {journal} {Communications in Mathematical Physics}\ }\textbf {\bibinfo
  {volume} {346}},\ \bibinfo {pages} {397} (\bibinfo {year}
  {2016})}\BibitemShut {NoStop}%
\bibitem [{\citenamefont {Harrow}\ and\ \citenamefont
  {Mehraban}(2023)}]{Harrow2023approximate}%
  \BibitemOpen
  \bibfield  {author} {\bibinfo {author} {\bibfnamefont {A.~W.}\ \bibnamefont
  {Harrow}}\ and\ \bibinfo {author} {\bibfnamefont {S.}~\bibnamefont
  {Mehraban}},\ }\bibfield  {journal} {\bibinfo  {journal} {Communications in
  Mathematical Physics}\ }\href {https://doi.org/10.1007/s00220-023-04675-z}
  {10.1007/s00220-023-04675-z} (\bibinfo {year} {2023})\BibitemShut {NoStop}%
\bibitem [{\citenamefont {Hunter-Jones}(2019)}]{hunterjones2019unitary}%
  \BibitemOpen
  \bibfield  {author} {\bibinfo {author} {\bibfnamefont {N.}~\bibnamefont
  {Hunter-Jones}},\ }\href@noop {} {\bibinfo {title} {Unitary designs from
  statistical mechanics in random quantum circuits}} (\bibinfo {year} {2019}),\
  \Eprint {https://arxiv.org/abs/1905.12053} {arXiv:1905.12053 [quant-ph]}
  \BibitemShut {NoStop}%
\bibitem [{\citenamefont {Lashkari}\ \emph {et~al.}(2013)\citenamefont
  {Lashkari}, \citenamefont {Stanford}, \citenamefont {Hastings}, \citenamefont
  {Osborne},\ and\ \citenamefont {Hayden}}]{Lashkari_2013}%
  \BibitemOpen
  \bibfield  {author} {\bibinfo {author} {\bibfnamefont {N.}~\bibnamefont
  {Lashkari}}, \bibinfo {author} {\bibfnamefont {D.}~\bibnamefont {Stanford}},
  \bibinfo {author} {\bibfnamefont {M.}~\bibnamefont {Hastings}}, \bibinfo
  {author} {\bibfnamefont {T.}~\bibnamefont {Osborne}},\ and\ \bibinfo {author}
  {\bibfnamefont {P.}~\bibnamefont {Hayden}},\ }\bibfield  {journal} {\bibinfo
  {journal} {Journal of High Energy Physics}\ }\textbf {\bibinfo {volume}
  {2013}},\ \href {https://doi.org/10.1007/jhep04(2013)022}
  {10.1007/jhep04(2013)022} (\bibinfo {year} {2013})\BibitemShut {NoStop}%
\bibitem [{\citenamefont {Onorati}\ \emph {et~al.}(2017)\citenamefont
  {Onorati}, \citenamefont {Buerschaper}, \citenamefont {Kliesch},
  \citenamefont {Brown}, \citenamefont {Werner},\ and\ \citenamefont
  {Eisert}}]{Onorati_2017}%
  \BibitemOpen
  \bibfield  {author} {\bibinfo {author} {\bibfnamefont {E.}~\bibnamefont
  {Onorati}}, \bibinfo {author} {\bibfnamefont {O.}~\bibnamefont
  {Buerschaper}}, \bibinfo {author} {\bibfnamefont {M.}~\bibnamefont
  {Kliesch}}, \bibinfo {author} {\bibfnamefont {W.}~\bibnamefont {Brown}},
  \bibinfo {author} {\bibfnamefont {A.~H.}\ \bibnamefont {Werner}},\ and\
  \bibinfo {author} {\bibfnamefont {J.}~\bibnamefont {Eisert}},\ }\href
  {https://doi.org/10.1007/s00220-017-2950-6} {\bibfield  {journal} {\bibinfo
  {journal} {Communications in Mathematical Physics}\ }\textbf {\bibinfo
  {volume} {355}},\ \bibinfo {pages} {905} (\bibinfo {year}
  {2017})}\BibitemShut {NoStop}%
\bibitem [{\citenamefont {Bentsen}\ \emph {et~al.}(2021)\citenamefont
  {Bentsen}, \citenamefont {Sahu},\ and\ \citenamefont
  {Swingle}}]{Bentsen_2021}%
  \BibitemOpen
  \bibfield  {author} {\bibinfo {author} {\bibfnamefont {G.~S.}\ \bibnamefont
  {Bentsen}}, \bibinfo {author} {\bibfnamefont {S.}~\bibnamefont {Sahu}},\ and\
  \bibinfo {author} {\bibfnamefont {B.}~\bibnamefont {Swingle}},\ }\bibfield
  {journal} {\bibinfo  {journal} {Physical Review B}\ }\textbf {\bibinfo
  {volume} {104}},\ \href {https://doi.org/10.1103/physrevb.104.094304}
  {10.1103/physrevb.104.094304} (\bibinfo {year} {2021})\BibitemShut {NoStop}%
\bibitem [{\citenamefont {Sahu}\ \emph {et~al.}(2022)\citenamefont {Sahu},
  \citenamefont {Jian}, \citenamefont {Bentsen},\ and\ \citenamefont
  {Swingle}}]{Sahu_2022}%
  \BibitemOpen
  \bibfield  {author} {\bibinfo {author} {\bibfnamefont {S.}~\bibnamefont
  {Sahu}}, \bibinfo {author} {\bibfnamefont {S.-K.}\ \bibnamefont {Jian}},
  \bibinfo {author} {\bibfnamefont {G.}~\bibnamefont {Bentsen}},\ and\ \bibinfo
  {author} {\bibfnamefont {B.}~\bibnamefont {Swingle}},\ }\bibfield  {journal}
  {\bibinfo  {journal} {Physical Review B}\ }\textbf {\bibinfo {volume}
  {106}},\ \href {https://doi.org/10.1103/physrevb.106.224305}
  {10.1103/physrevb.106.224305} (\bibinfo {year} {2022})\BibitemShut {NoStop}%
\bibitem [{\citenamefont {Jian}\ \emph {et~al.}(2022)\citenamefont {Jian},
  \citenamefont {Bentsen},\ and\ \citenamefont {Swingle}}]{jian2022linear}%
  \BibitemOpen
  \bibfield  {author} {\bibinfo {author} {\bibfnamefont {S.-K.}\ \bibnamefont
  {Jian}}, \bibinfo {author} {\bibfnamefont {G.}~\bibnamefont {Bentsen}},\ and\
  \bibinfo {author} {\bibfnamefont {B.}~\bibnamefont {Swingle}},\ }\href@noop
  {} {\bibfield  {journal} {\bibinfo  {journal} {arXiv preprint
  arXiv:2206.14205}\ } (\bibinfo {year} {2022})}\BibitemShut {NoStop}%
\bibitem [{\citenamefont {Vidal}(2003)}]{Vidal_2003}%
  \BibitemOpen
  \bibfield  {author} {\bibinfo {author} {\bibfnamefont {G.}~\bibnamefont
  {Vidal}},\ }\href {https://doi.org/10.1103/PhysRevLett.91.147902} {\bibfield
  {journal} {\bibinfo  {journal} {Phys. Rev. Lett.}\ }\textbf {\bibinfo
  {volume} {91}},\ \bibinfo {pages} {147902} (\bibinfo {year}
  {2003})}\BibitemShut {NoStop}%
\bibitem [{\citenamefont {White}\ and\ \citenamefont
  {Feiguin}(2004)}]{White_2004}%
  \BibitemOpen
  \bibfield  {author} {\bibinfo {author} {\bibfnamefont {S.~R.}\ \bibnamefont
  {White}}\ and\ \bibinfo {author} {\bibfnamefont {A.~E.}\ \bibnamefont
  {Feiguin}},\ }\href {https://doi.org/10.1103/PhysRevLett.93.076401}
  {\bibfield  {journal} {\bibinfo  {journal} {Phys. Rev. Lett.}\ }\textbf
  {\bibinfo {volume} {93}},\ \bibinfo {pages} {076401} (\bibinfo {year}
  {2004})}\BibitemShut {NoStop}%
\bibitem [{\citenamefont {Daley}\ \emph {et~al.}(2004)\citenamefont {Daley},
  \citenamefont {Kollath}, \citenamefont {Schollwöck},\ and\ \citenamefont
  {Vidal}}]{Daley_2004}%
  \BibitemOpen
  \bibfield  {author} {\bibinfo {author} {\bibfnamefont {A.~J.}\ \bibnamefont
  {Daley}}, \bibinfo {author} {\bibfnamefont {C.}~\bibnamefont {Kollath}},
  \bibinfo {author} {\bibfnamefont {U.}~\bibnamefont {Schollwöck}},\ and\
  \bibinfo {author} {\bibfnamefont {G.}~\bibnamefont {Vidal}},\ }\href
  {https://doi.org/10.1088/1742-5468/2004/04/p04005} {\bibfield  {journal}
  {\bibinfo  {journal} {Journal of Statistical Mechanics: Theory and
  Experiment}\ }\textbf {\bibinfo {volume} {2004}},\ \bibinfo {pages} {P04005}
  (\bibinfo {year} {2004})}\BibitemShut {NoStop}%
\bibitem [{\citenamefont {Jian}\ and\ \citenamefont
  {Swingle}(2021)}]{jian2021phase}%
  \BibitemOpen
  \bibfield  {author} {\bibinfo {author} {\bibfnamefont {S.-K.}\ \bibnamefont
  {Jian}}\ and\ \bibinfo {author} {\bibfnamefont {B.}~\bibnamefont {Swingle}},\
  }\href@noop {} {\bibfield  {journal} {\bibinfo  {journal} {arXiv preprint
  arXiv:2108.11973}\ } (\bibinfo {year} {2021})}\BibitemShut {NoStop}%
\bibitem [{\citenamefont {Napp}\ \emph {et~al.}(2022)\citenamefont {Napp},
  \citenamefont {Placa}, \citenamefont {Dalzell}, \citenamefont
  {Brand{\~{a}}o},\ and\ \citenamefont {Harrow}}]{Napp_2022}%
  \BibitemOpen
  \bibfield  {author} {\bibinfo {author} {\bibfnamefont {J.~C.}\ \bibnamefont
  {Napp}}, \bibinfo {author} {\bibfnamefont {R.~L.~L.}\ \bibnamefont {Placa}},
  \bibinfo {author} {\bibfnamefont {A.~M.}\ \bibnamefont {Dalzell}}, \bibinfo
  {author} {\bibfnamefont {F.~G.}\ \bibnamefont {Brand{\~{a}}o}},\ and\
  \bibinfo {author} {\bibfnamefont {A.~W.}\ \bibnamefont {Harrow}},\ }\bibfield
   {journal} {\bibinfo  {journal} {Physical Review X}\ }\textbf {\bibinfo
  {volume} {12}},\ \href {https://doi.org/10.1103/physrevx.12.021021}
  {10.1103/physrevx.12.021021} (\bibinfo {year} {2022})\BibitemShut {NoStop}%
\bibitem [{\citenamefont {Balasubramanian}\ \emph {et~al.}(2023)\citenamefont
  {Balasubramanian}, \citenamefont {Kar}, \citenamefont {Li}, \citenamefont
  {Parrikar},\ and\ \citenamefont {Rajgadia}}]{balasubramanian2023quantum}%
  \BibitemOpen
  \bibfield  {author} {\bibinfo {author} {\bibfnamefont {V.}~\bibnamefont
  {Balasubramanian}}, \bibinfo {author} {\bibfnamefont {A.}~\bibnamefont
  {Kar}}, \bibinfo {author} {\bibfnamefont {C.}~\bibnamefont {Li}}, \bibinfo
  {author} {\bibfnamefont {O.}~\bibnamefont {Parrikar}},\ and\ \bibinfo
  {author} {\bibfnamefont {H.}~\bibnamefont {Rajgadia}},\ }\href@noop {}
  {\bibinfo {title} {Quantum error correction from complexity in brownian syk}}
  (\bibinfo {year} {2023}),\ \Eprint {https://arxiv.org/abs/2301.07108}
  {arXiv:2301.07108 [hep-th]} \BibitemShut {NoStop}%
\bibitem [{\citenamefont {Ware}\ \emph {et~al.}(2023)\citenamefont {Ware},
  \citenamefont {Deshpande}, \citenamefont {Hangleiter}, \citenamefont
  {Niroula}, \citenamefont {Fefferman}, \citenamefont {Gorshkov},\ and\
  \citenamefont {Gullans}}]{ware2023sharp}%
  \BibitemOpen
  \bibfield  {author} {\bibinfo {author} {\bibfnamefont {B.}~\bibnamefont
  {Ware}}, \bibinfo {author} {\bibfnamefont {A.}~\bibnamefont {Deshpande}},
  \bibinfo {author} {\bibfnamefont {D.}~\bibnamefont {Hangleiter}}, \bibinfo
  {author} {\bibfnamefont {P.}~\bibnamefont {Niroula}}, \bibinfo {author}
  {\bibfnamefont {B.}~\bibnamefont {Fefferman}}, \bibinfo {author}
  {\bibfnamefont {A.~V.}\ \bibnamefont {Gorshkov}},\ and\ \bibinfo {author}
  {\bibfnamefont {M.~J.}\ \bibnamefont {Gullans}},\ }\href@noop {} {\bibinfo
  {title} {A sharp phase transition in linear cross-entropy benchmarking}}
  (\bibinfo {year} {2023}),\ \Eprint {https://arxiv.org/abs/2305.04954}
  {arXiv:2305.04954 [quant-ph]} \BibitemShut {NoStop}%
\bibitem [{\citenamefont {Dalzell}\ \emph {et~al.}(2021)\citenamefont
  {Dalzell}, \citenamefont {Hunter-Jones},\ and\ \citenamefont
  {Brandão}}]{dalzell2021random}%
  \BibitemOpen
  \bibfield  {author} {\bibinfo {author} {\bibfnamefont {A.~M.}\ \bibnamefont
  {Dalzell}}, \bibinfo {author} {\bibfnamefont {N.}~\bibnamefont
  {Hunter-Jones}},\ and\ \bibinfo {author} {\bibfnamefont {F.~G. S.~L.}\
  \bibnamefont {Brandão}},\ }\href@noop {} {\bibinfo {title} {Random quantum
  circuits transform local noise into global white noise}} (\bibinfo {year}
  {2021}),\ \Eprint {https://arxiv.org/abs/2111.14907} {arXiv:2111.14907
  [quant-ph]} \BibitemShut {NoStop}%
\bibitem [{\citenamefont {Aharonov}\ \emph {et~al.}(2023)\citenamefont
  {Aharonov}, \citenamefont {Gao}, \citenamefont {Landau}, \citenamefont
  {Liu},\ and\ \citenamefont {Vazirani}}]{Aharonov_2023}%
  \BibitemOpen
  \bibfield  {author} {\bibinfo {author} {\bibfnamefont {D.}~\bibnamefont
  {Aharonov}}, \bibinfo {author} {\bibfnamefont {X.}~\bibnamefont {Gao}},
  \bibinfo {author} {\bibfnamefont {Z.}~\bibnamefont {Landau}}, \bibinfo
  {author} {\bibfnamefont {Y.}~\bibnamefont {Liu}},\ and\ \bibinfo {author}
  {\bibfnamefont {U.}~\bibnamefont {Vazirani}},\ }in\ \href
  {https://doi.org/10.1145/3564246.3585234} {\emph {\bibinfo {booktitle}
  {Proceedings of the 55th Annual {ACM} Symposium on Theory of Computing}}}\
  (\bibinfo  {publisher} {{ACM}},\ \bibinfo {year} {2023})\BibitemShut
  {NoStop}%
\bibitem [{\citenamefont {Barak}\ \emph {et~al.}(2020)\citenamefont {Barak},
  \citenamefont {Chou},\ and\ \citenamefont {Gao}}]{barak2020spoofing}%
  \BibitemOpen
  \bibfield  {author} {\bibinfo {author} {\bibfnamefont {B.}~\bibnamefont
  {Barak}}, \bibinfo {author} {\bibfnamefont {C.-N.}\ \bibnamefont {Chou}},\
  and\ \bibinfo {author} {\bibfnamefont {X.}~\bibnamefont {Gao}},\ }\href@noop
  {} {\bibinfo {title} {Spoofing linear cross-entropy benchmarking in shallow
  quantum circuits}} (\bibinfo {year} {2020}),\ \Eprint
  {https://arxiv.org/abs/2005.02421} {arXiv:2005.02421 [quant-ph]} \BibitemShut
  {NoStop}%
\bibitem [{\citenamefont {Gao}\ \emph {et~al.}(2021)\citenamefont {Gao},
  \citenamefont {Kalinowski}, \citenamefont {Chou}, \citenamefont {Lukin},
  \citenamefont {Barak},\ and\ \citenamefont {Choi}}]{gao2021limitations}%
  \BibitemOpen
  \bibfield  {author} {\bibinfo {author} {\bibfnamefont {X.}~\bibnamefont
  {Gao}}, \bibinfo {author} {\bibfnamefont {M.}~\bibnamefont {Kalinowski}},
  \bibinfo {author} {\bibfnamefont {C.-N.}\ \bibnamefont {Chou}}, \bibinfo
  {author} {\bibfnamefont {M.~D.}\ \bibnamefont {Lukin}}, \bibinfo {author}
  {\bibfnamefont {B.}~\bibnamefont {Barak}},\ and\ \bibinfo {author}
  {\bibfnamefont {S.}~\bibnamefont {Choi}},\ }\href@noop {} {\bibinfo {title}
  {Limitations of linear cross-entropy as a measure for quantum advantage}}
  (\bibinfo {year} {2021}),\ \Eprint {https://arxiv.org/abs/2112.01657}
  {arXiv:2112.01657 [quant-ph]} \BibitemShut {NoStop}%
\bibitem [{\citenamefont {Jian}\ \emph
  {et~al.}(2021{\natexlab{a}})\citenamefont {Jian}, \citenamefont {Liu},
  \citenamefont {Chen}, \citenamefont {Swingle},\ and\ \citenamefont
  {Zhang}}]{jian2021quantum}%
  \BibitemOpen
  \bibfield  {author} {\bibinfo {author} {\bibfnamefont {S.-K.}\ \bibnamefont
  {Jian}}, \bibinfo {author} {\bibfnamefont {C.}~\bibnamefont {Liu}}, \bibinfo
  {author} {\bibfnamefont {X.}~\bibnamefont {Chen}}, \bibinfo {author}
  {\bibfnamefont {B.}~\bibnamefont {Swingle}},\ and\ \bibinfo {author}
  {\bibfnamefont {P.}~\bibnamefont {Zhang}},\ }\href@noop {} {\bibfield
  {journal} {\bibinfo  {journal} {arXiv preprint arXiv:2106.09635}\ } (\bibinfo
  {year} {2021}{\natexlab{a}})}\BibitemShut {NoStop}%
\bibitem [{Note1()}]{Note1}%
  \BibitemOpen
  \bibinfo {note} {In a one-dimensional chain with local couplings, the
  elementary excitation manifests as a domain wall, with a finite gap
  independent of the system size and an entropy proportional to the system
  size}\BibitemShut {NoStop}%
\bibitem [{\citenamefont {Schumacher}\ and\ \citenamefont
  {Nielsen}(1996)}]{Schumacher_1996}%
  \BibitemOpen
  \bibfield  {author} {\bibinfo {author} {\bibfnamefont {B.}~\bibnamefont
  {Schumacher}}\ and\ \bibinfo {author} {\bibfnamefont {M.~A.}\ \bibnamefont
  {Nielsen}},\ }\href {https://doi.org/10.1103/physreva.54.2629} {\bibfield
  {journal} {\bibinfo  {journal} {Physical Review A}\ }\textbf {\bibinfo
  {volume} {54}},\ \bibinfo {pages} {2629} (\bibinfo {year}
  {1996})}\BibitemShut {NoStop}%
\bibitem [{\citenamefont {Schumacher}\ and\ \citenamefont
  {Westmoreland}(2001)}]{schumacher2001approximate}%
  \BibitemOpen
  \bibfield  {author} {\bibinfo {author} {\bibfnamefont {B.}~\bibnamefont
  {Schumacher}}\ and\ \bibinfo {author} {\bibfnamefont {M.~D.}\ \bibnamefont
  {Westmoreland}},\ }\href@noop {} {\bibinfo {title} {Approximate quantum error
  correction}} (\bibinfo {year} {2001}),\ \Eprint
  {https://arxiv.org/abs/quant-ph/0112106} {arXiv:quant-ph/0112106 [quant-ph]}
  \BibitemShut {NoStop}%
\bibitem [{\citenamefont {Choi}(1975)}]{choi_completely_1975}%
  \BibitemOpen
  \bibfield  {author} {\bibinfo {author} {\bibfnamefont {M.-D.}\ \bibnamefont
  {Choi}},\ }\href
  {https://doi.org/https://doi.org/10.1016/0024-3795(75)90075-0} {\bibfield
  {journal} {\bibinfo  {journal} {Linear Algebra and its Applications}\
  }\textbf {\bibinfo {volume} {10}},\ \bibinfo {pages} {285} (\bibinfo {year}
  {1975})}\BibitemShut {NoStop}%
\bibitem [{\citenamefont {Jamiołkowski}(1972)}]{jamiolkowski}%
  \BibitemOpen
  \bibfield  {author} {\bibinfo {author} {\bibfnamefont {A.}~\bibnamefont
  {Jamiołkowski}},\ }\href
  {https://doi.org/https://doi.org/10.1016/0034-4877(72)90011-0} {\bibfield
  {journal} {\bibinfo  {journal} {Reports on Mathematical Physics}\ }\textbf
  {\bibinfo {volume} {3}},\ \bibinfo {pages} {275} (\bibinfo {year}
  {1972})}\BibitemShut {NoStop}%
\bibitem [{\citenamefont {Hauschild}\ and\ \citenamefont
  {Pollmann}(2018)}]{tenpy}%
  \BibitemOpen
  \bibfield  {author} {\bibinfo {author} {\bibfnamefont {J.}~\bibnamefont
  {Hauschild}}\ and\ \bibinfo {author} {\bibfnamefont {F.}~\bibnamefont
  {Pollmann}},\ }\href {https://doi.org/10.21468/SciPostPhysLectNotes.5}
  {\bibfield  {journal} {\bibinfo  {journal} {SciPost Phys. Lect. Notes}\ ,\
  \bibinfo {pages} {5}} (\bibinfo {year} {2018})},\ \bibinfo {note} {code
  available from \url{https://github.com/tenpy/tenpy}},\ \Eprint
  {https://arxiv.org/abs/1805.00055} {arXiv:1805.00055} \BibitemShut {NoStop}%
\bibitem [{Note2()}]{Note2}%
  \BibitemOpen
  \bibinfo {note} {Note that since the variance of coupling is independent of
  $\alpha = x,y,z$, the resulted Hamiltonian also enjoys a $SU(2)$ symmetry for
  each site. But our results do not rely on this symmetry.}\BibitemShut {Stop}%
\bibitem [{\citenamefont {Binder}\ and\ \citenamefont
  {Landau}(1984)}]{binder1984finite}%
  \BibitemOpen
  \bibfield  {author} {\bibinfo {author} {\bibfnamefont {K.}~\bibnamefont
  {Binder}}\ and\ \bibinfo {author} {\bibfnamefont {D.}~\bibnamefont
  {Landau}},\ }\href@noop {} {\bibfield  {journal} {\bibinfo  {journal}
  {Physical Review B}\ }\textbf {\bibinfo {volume} {30}},\ \bibinfo {pages}
  {1477} (\bibinfo {year} {1984})}\BibitemShut {NoStop}%
\bibitem [{\citenamefont {Brandao}\ and\ \citenamefont
  {Kastoryano}(2019)}]{brandao2019finite}%
  \BibitemOpen
  \bibfield  {author} {\bibinfo {author} {\bibfnamefont {F.~G. S.~L.}\
  \bibnamefont {Brandao}}\ and\ \bibinfo {author} {\bibfnamefont {M.~J.}\
  \bibnamefont {Kastoryano}},\ }\href@noop {} {\bibinfo {title} {Finite
  correlation length implies efficient preparation of quantum thermal states}}
  (\bibinfo {year} {2019}),\ \Eprint {https://arxiv.org/abs/1609.07877}
  {arXiv:1609.07877 [quant-ph]} \BibitemShut {NoStop}%
\bibitem [{Note3()}]{Note3}%
  \BibitemOpen
  \bibinfo {note} {For a more accurate estimate, we need to rescale $N$ by a
  factor $c_3 < 1$. This is because spin flips do not interact with each other
  only when they are dilute enough.}\BibitemShut {Stop}%
\bibitem [{\citenamefont {Abeyesinghe}\ \emph {et~al.}(2009)\citenamefont
  {Abeyesinghe}, \citenamefont {Devetak}, \citenamefont {Hayden},\ and\
  \citenamefont {Winter}}]{Abeyesinghe_2009}%
  \BibitemOpen
  \bibfield  {author} {\bibinfo {author} {\bibfnamefont {A.}~\bibnamefont
  {Abeyesinghe}}, \bibinfo {author} {\bibfnamefont {I.}~\bibnamefont
  {Devetak}}, \bibinfo {author} {\bibfnamefont {P.}~\bibnamefont {Hayden}},\
  and\ \bibinfo {author} {\bibfnamefont {A.}~\bibnamefont {Winter}},\ }\href
  {https://doi.org/10.1098/rspa.2009.0202} {\bibfield  {journal} {\bibinfo
  {journal} {Proceedings of the Royal Society A: Mathematical, Physical and
  Engineering Sciences}\ }\textbf {\bibinfo {volume} {465}},\ \bibinfo {pages}
  {2537} (\bibinfo {year} {2009})}\BibitemShut {NoStop}%
\bibitem [{\citenamefont {B\'eny}\ and\ \citenamefont
  {Oreshkov}(2010)}]{Beny_General_2010}%
  \BibitemOpen
  \bibfield  {author} {\bibinfo {author} {\bibfnamefont {C.}~\bibnamefont
  {B\'eny}}\ and\ \bibinfo {author} {\bibfnamefont {O.}~\bibnamefont
  {Oreshkov}},\ }\href {https://doi.org/10.1103/PhysRevLett.104.120501}
  {\bibfield  {journal} {\bibinfo  {journal} {Phys. Rev. Lett.}\ }\textbf
  {\bibinfo {volume} {104}},\ \bibinfo {pages} {120501} (\bibinfo {year}
  {2010})}\BibitemShut {NoStop}%
\bibitem [{Note4()}]{Note4}%
  \BibitemOpen
  \bibinfo {note} {Note that in the symmetry breaking phase, there are still
  two phases for domain walls, the pinning and depinning phases, separating by
  a pinning transition. For our case, since the coupling at the top layer is
  the same as the coupling in the bulk, the depinning transition is the same as
  the symmetry breaking transition. This means the domain wall is depinned and
  can fluctuate.}\BibitemShut {Stop}%
\bibitem [{\citenamefont {Abraham}(1980)}]{abraham1980solvable}%
  \BibitemOpen
  \bibfield  {author} {\bibinfo {author} {\bibfnamefont {D.}~\bibnamefont
  {Abraham}},\ }\href@noop {} {\bibfield  {journal} {\bibinfo  {journal}
  {Physical Review Letters}\ }\textbf {\bibinfo {volume} {44}},\ \bibinfo
  {pages} {1165} (\bibinfo {year} {1980})}\BibitemShut {NoStop}%
\bibitem [{\citenamefont {Jian}\ \emph
  {et~al.}(2021{\natexlab{b}})\citenamefont {Jian}, \citenamefont {Liu},
  \citenamefont {Chen}, \citenamefont {Swingle},\ and\ \citenamefont
  {Zhang}}]{jian2021measurement}%
  \BibitemOpen
  \bibfield  {author} {\bibinfo {author} {\bibfnamefont {S.-K.}\ \bibnamefont
  {Jian}}, \bibinfo {author} {\bibfnamefont {C.}~\bibnamefont {Liu}}, \bibinfo
  {author} {\bibfnamefont {X.}~\bibnamefont {Chen}}, \bibinfo {author}
  {\bibfnamefont {B.}~\bibnamefont {Swingle}},\ and\ \bibinfo {author}
  {\bibfnamefont {P.}~\bibnamefont {Zhang}},\ }\href@noop {} {\bibfield
  {journal} {\bibinfo  {journal} {Physical review letters}\ }\textbf {\bibinfo
  {volume} {127}},\ \bibinfo {pages} {140601} (\bibinfo {year}
  {2021}{\natexlab{b}})}\BibitemShut {NoStop}%
\bibitem [{Note5()}]{Note5}%
  \BibitemOpen
  \bibinfo {note} {See Appendix. B in~\cite
  {balasubramanian2023quantum}}\BibitemShut {NoStop}%
\end{thebibliography}%

\newpage
\onecolumngrid

\setcounter{secnumdepth}{3}
\setcounter{equation}{0}
\setcounter{figure}{0}
\renewcommand{\theequation}{S\arabic{equation}}
\renewcommand{\thefigure}{S\arabic{figure}}
\renewcommand\figurename{Supplementary Figure}
\renewcommand\tablename{Supplementary Table}
\newcommand\Scite[1]{[S\citealp{#1}]}
\makeatletter \renewcommand\@biblabel[1]{[S#1]} \makeatother

\appendix

\section{Quantum Error Correcting codes generated by Brownian circuits}\label{appsec:qec}
\subsection{Error correction bound with mutual purity}\label{appsec:qec_proof}
In this section we will briefly recap the derivation of the error correction bound Eq.~\ref{eq:approx_qec2} derived in~\cite{balasubramanian2023quantum}. We will assume the setup described in Fig.~\ref{fig:qecc_decoupling}. Consider first the encoding of the maximally entangled state between the reference $R$ and system $A$,
\bea 
\ket{\Psi}_{RA} = \frac{1}{\sqrt{d_{R}}}\sum_{i = 1}^{d_{R}}\ket{i}_{R}\ket{\phi_{i}}_{A},
\eea
before any application of error. After the error channel acts on $A$, we get the following noise-affected state on $RAE$,
\bea 
&\ket{\Psi^{\prime}}_{RAE} =  \frac{1}{\sqrt{d_{R}}}\sum_{i = 1}^{d_{R}}\ket{i}_{R}U_{\text{err}}\left(\ket{\phi_{i}}_{A}\ket{e_{0}}_{E}\right) \nn.
\eea

The error recovery procedure $\mathcal{R}$~\cite{schumacher2001approximate,balasubramanian2023quantum} works by first measuring $A$ using an ideal projective measurement that probes the effect of error in $\ket{\Psi^{\prime}}_{RAE}$, followed by a unitary update of the state to restore it to $\ket{\Psi}_{RA}$. We first introduce an an orthonormal basis of states in $A$, $\ket{\phi_{ij}}_{A}$.  The projective measurement is given by the projection operators,
\bea
\Pi_{j} = \sum_{i=1}^{d_{R}} \ket{\phi_{ij}}_{A}\bra{\phi_{ij}}_{A}.
\eea
Depending on the measurement outcome, a corrective unitary $U_{j, A}$ is applied on system $A$. In order to study the effectiveness of the recovery channel, we want to study the trace distance between the recovered state and the encoded state, $\bigg|\bigg|\mathcal{R}\left(\ket{\Psi^{\prime}}\bra{\Psi^{\prime}}\right),\ket{\Psi}\bra{\Psi}\bigg|\bigg|_{1}$. In order to bound this, we introduce a fictitious state $\tilde{\ket{\Psi}}_{RAE}$ which aids in the analysis.

Consider $\tilde{\rho}_{RE} = \rho^{\prime}_{R} \otimes \rho^{\prime}_{E}$, where the reduced density matrices $\rho^{\prime}$ are obtained from the state $\ket{\Psi^{\prime}}_{RAE}$. We now take the fictitious state $\tilde{\ket{\Psi}}_{RAE}$ which is a purification of $\tilde{\rho}_{RE}$ such that the trace distance between $\tilde{\ket{\Psi}}_{RAE}$ and $\ket{\Psi^{\prime}}_{RAE}$ is minimal. This uniquely defines,
\bea 
&\tilde{\ket{\Psi}}_{RAE} =  \frac{1}{\sqrt{d_{R}}}\sum_{i = 1}^{d_{R}}\sum_{j = 1}^{d_{E}}
\sqrt{\alpha_{j}}
\ket{i}_{R}
\ket{\phi_{ij}}_{A}
\ket{e_{j}}_{E}.
\eea

Imagine we apply the recovery channel $\mathcal{R}$ on $\tilde{\ket{\Psi}}_{RAE}$ instead. After the measurement, say the outcome $j$ is obtained. The measured fictitious state is now,
\bea 
&\tilde{\ket{\Psi}}_{RAE}^{j} =  \frac{1}{\sqrt{d_{R}}}\sum_{i = 1}^{d_{R}}\ket{i}_{R}\ket{\phi_{ij}}_{A}\ket{e_{j}}_{E}.
\eea
We can now choose $U_{j}$ acting on $A$ such that $U_{j,A}\ket{\phi_{ij}}_{A} = \ket{\phi_{i}}_{A}$, and we get,
\bea
U_{j,A}\tilde{\ket{\Psi}}_{RAE}^{j} = \ket{\Psi}_{RA}\ket{e_{j}}_{E}.
\eea

From the above relation, we find that,
\begin{align}
\bigg|\bigg|\mathcal{R}\left(\ket{\Psi^{\prime}}\bra{\Psi^{\prime}}\right),\ket{\Psi}\bra{\Psi}\bigg|\bigg|_{1} &= \bigg|\bigg|\mathcal{R}\left(\ket{\Psi^{\prime}}\bra{\Psi^{\prime}}\right),\mathcal{R}\left(\tilde{\ket{\Psi}}\tilde{\bra{\Psi}}\right)\bigg|\bigg|_{1}\leq \bigg|\bigg|\ket{\Psi^{\prime}}\bra{\Psi^{\prime}},\tilde{\ket{\Psi}}\tilde{\bra{\Psi}}\bigg|\bigg|_{1}.
\end{align}
where the last inequality follows from the monotonicity property of the trace distance. In~\cite{balasubramanian2023quantum}, the last expression is bounded by the Mutual Purity defined in Eq.~\ref{eq:mutual_purity_exp}\footnote{See Appendix. B in~\cite{balasubramanian2023quantum}}. We quote the result in Eq.~\ref{eq:approx_qec2}.

\subsection{Replica computation of mutual purity}\label{appsec:qec_mp}

We first represent the reduced density matrix of the noise-affected state defined in Eq.~\ref{eq:noisy_state},
\bea
\rho^{\prime}_{RE} = \Tr_{A}\ket{\Psi^{\prime}}\bra{\Psi^{\prime}} = \frac{1}{d_{R}}\sum_{i,j = 1}^{d_{R}}\ket{i}\bra{j}_{R}\otimes \Tr_{A} \left\{U_{\text{err}}\left(U_{\text{enc}}\left(\ket{i}\bra{j}_{A_{1}}\otimes \ket{0}\bra{0}_{A_{2}}\right)U_{\text{enc}}^{\dagger}\otimes \ket{e_{0}}\bra{e_{0}}\right)U_{\text{err}}^{\dagger}\right\}
\eea

The effect of the $U_{\text{err}}$ on the system and the environment can be represented by Kraus operators acting on the system itself, 
\bea
U_{\text{err}}\ket{\psi}_{A}\ket{e_{0}}_{E} = \sum_{m} E^{m}_{A} \ket{\psi}_{A}\ket{e_{m}}_{E}\\E_{A}^{m}: \mathcal{H}_{A} \to \mathcal{H}_{A}, \quad  \sum_{m}E^{m\dagger}_{A}E^{m}_{A} = \mathbf{1}_{A}. \nonumber
\eea

The squared density matrix 
$\rho^{\prime\otimes 2}_{RE}$ can be represented by a state vector in the replicated Hilbert space $\mathcal{H}\otimes \mathcal{H^{*}}\otimes \mathcal{H}\otimes \mathcal{H^{*}}$, and the replicated unitaries $\mathbb{U}_{\text{enc}} = U_{\text{enc}}\otimes U^{*}_{\text{enc}}\otimes U_{\text{enc}}\otimes U^{*}_{\text{enc}}$ and $\mathbb{U}_{\text{err}} = U_{\text{err}}\otimes U^{*}_{\text{err}}\otimes U_{\text{err}}\otimes U^{*}_{\text{err}}$ as follows,
\bea
|\rho^{\prime \otimes 2}_{RE}\rangle \rangle = \frac{1}{d_{R}^{2}} \sum_{i,j,i^{\prime},j^{\prime}=1}^{d_{R}}|iji^{\prime}j^{\prime}\rangle\rangle_{R} \sum_{s,k = 1}^{d_{A}} \langle \langle sskk|_{A} \mathbb{U}_{\text{err}}\otimes \mathbb{U}_{\text{enc}}|iji^{\prime}j^{\prime}\rangle \rangle_{A_{1}}|0^{\otimes 4|A_{2}|}\rangle \rangle_{A_{2}}|e_{0}^{\otimes 4}\rangle\rangle_{E}.
\label{eq:double_rho}
\eea

While this representation looks cumbersome, it makes further computations straightforward. The mutual purity is given by $\mathcal{F}^{\Psi^{\prime}}_{RE} = \Tr \rho_{RE}^{\prime 2}-\Tr \rho_{R}^{\prime 2}\Tr \rho_{E}^{\prime 2}$. Let us compute each term. 

It is convenient to express Eq.~\ref{eq:double_rho} pictorially, with 4-rank tensors for each subsystem, representing $\mathcal{H}\otimes \mathcal{H^{*}}\otimes \mathcal{H}\otimes \mathcal{H^{*}}$,
\bea
\includegraphics[width = 0.35\columnwidth]{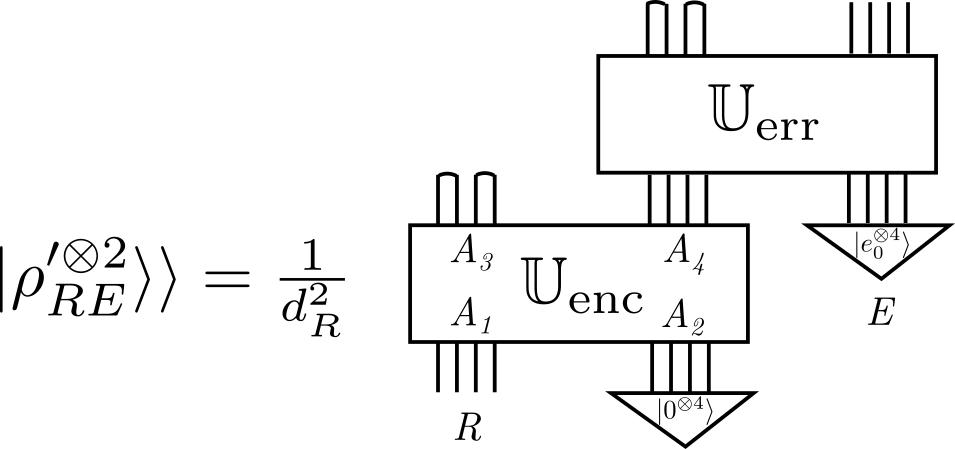} \nn
\eea

By unitarity of $U_{\text{err}}$ and $U_{\text{enc}}$ we have, 
$$\includegraphics[width = 0.1\columnwidth]{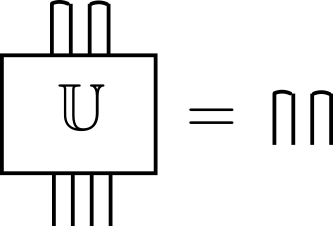}.$$

Using this result we find,
\bea
&&\includegraphics[width = 0.8\columnwidth]{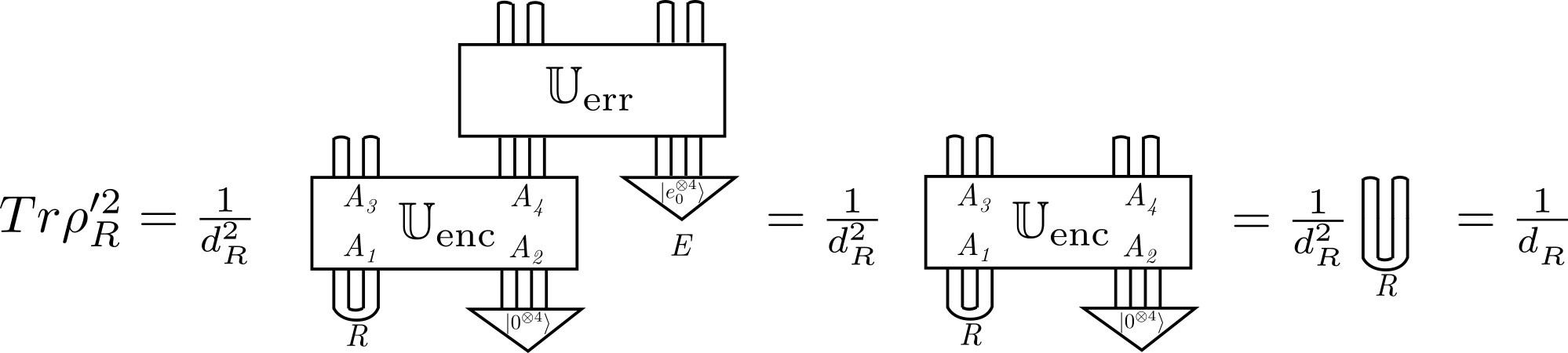} \nn \\
&&\includegraphics[width = 0.8\columnwidth]{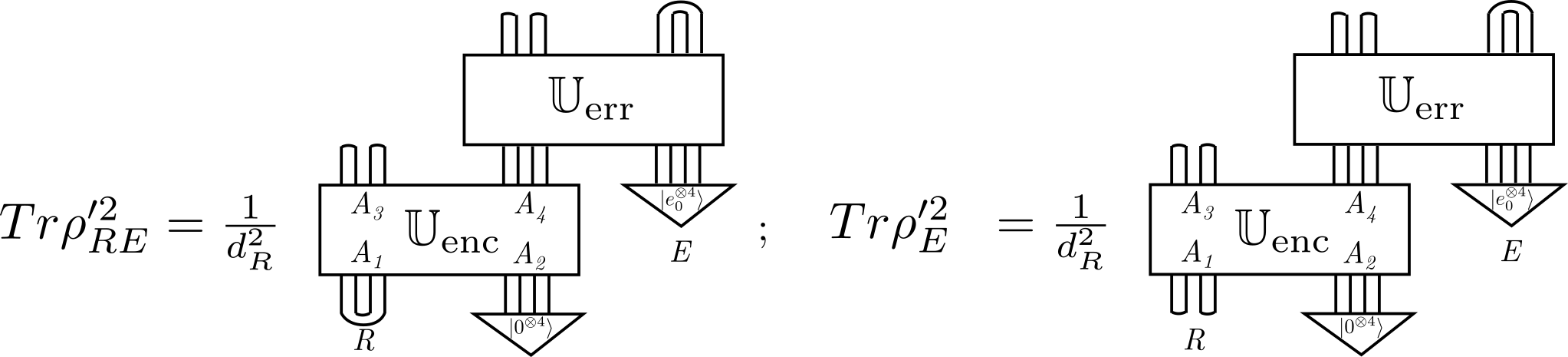} 
\eea

In general, $d_{R} = 2^{|A_{2}|}$, where  the local Hilbert space dimension is 2. We introduce notation for the following states in the replicated Hilbert space on $A$,
\begin{align}
    &|\psi_{\text{in}}\rangle \rangle = \left(\kettswap - \frac{1}{2} \kettid\right)^{\otimes A_{1}} \otimes\kett{0000}^{\otimes A_{2}} \\ 
    &|\psi_{\text{err}}\rangle \rangle = 2^{N}\sum_{m,n=0}^{d_{E}-1} E_{m}E_{n}^{*}E_{n}E_{m}^{*}\kettid^{\otimes A}.
\end{align}
Note $\kett{\psi_{\text{err}}}$ is an unnormalized state and the expression for $\kett{\psi_{\text{err}}}$ includes the case for no error acting on the subsystem $A_{3} \subset A$ by choosing $E_{m} = \mathbf{1}_{A_{3}}\otimes E_{m, A_{4}}$. 

Combining all the expressions, the mutual purity is given by,
\bea
\mathcal{F}^{\Psi^{\prime}}_{RE} = \langle\langle \psi_{\text{err}}|\mathbb{U}_{\text{enc}}|\psi_{\text{in}}\rangle \rangle.
\label{eq:mutual_purity_exp}
\eea

%This expression can be evaluated for $\mathbb{U}_{\text{enc}}$ generated by a Brownian circuit using the imaginary TEBD formalism.

The noise model for probing the extent of error correction enters the computation of mutual purity only in the definition of $\kett{\psi_{\text{err}}}$. Consider the local depolarization channel of strength $\lambda$ on a subset $A_{4}\subset A$, such that number of qubits undergoing noise is $|A_{4}| = p|A|$. The Kraus operators for the depolarization channel are,
\bea
E_{0} = \sqrt{1-\lambda} \mathbf{1}, E_{x,y,z} = \sqrt{\frac{\lambda}{3}}\sigma_{x,y,z}.
\eea
Local depolarization channel acting on each qubit can be purified using a environment degree of freedom with $4$ levels $0,x,y,z$. The corresponding $d_{\text{env}} = 4^{p|A|}$.

\subsection{Maximal complexity encoding by Haar random circuits}\label{appsec:qec_haar}

We can compute the mutual purity for any noise model for an encoding unitary $U_{\text{enc}}$ which is a global Haar random unitary. Any unitary 2 designs will exhibit this value of mutual purity. By examining the time it requires for the Brownian circuit to achieve this value of mutual purity, we can diagnose the time required for the Brownian circuit to realise a 2 design.

For a global Haar random unitary, we have,

\bea
U_{\text{Haar}}\otimes U^{*}_{\text{Haar}}\otimes U_{\text{Haar}}\otimes U^{*}_{\text{Haar}} = \frac{1}{d^{2}-1}\left(\kettid\braaid+\kettswap\braaswap-\frac{1}{d}\left(\kettid\braaswap+\kettswap\braaid\right)\right).
\eea
Using this identity for $U_{\text{enc}}$ in Eq.~\ref{eq:mutual_purity_exp}, we get the following terms in the expression for mutual purity,
\bea
    &\mathbb{E}[\Tr \rho_{RE}^2] = \left(\frac1{\sqrt{d_R}}\right)^4 \frac1{d_A^2-1} \left( f_\text{id}(\lambda) d_R  +  f_\text{swap}(\lambda) d_R^2  - \frac{1}{d_{A}}\left( f_\text{id}(\lambda) d_R^2 +   f_\text{swap}(\lambda) d_R \right)\right), \nn \\
    &\mathbb{E}[\Tr \rho_{E}^2 ] = \left(\frac1{\sqrt{d_R}}\right)^4 \frac1{d_A^2-1} \left( f_\text{id}(\lambda) d_R^2  + f_\text{swap}(\lambda) d_R  - \frac{1}{d_A}\left( f_\text{id}(\lambda) d_R +   f_\text{swap}(\lambda) d_R^2 \right)\right), \nn
\eea
where we have introduced the notation,
\bea
    f_\text{id}(\lambda) = \braa{\psi_{\text{err}}}\text{id}\rangle\rangle , \quad f_\text{swap}(\lambda) = \braa{\psi_{\text{err}}}\text{swap}\rangle\rangle \nn.
\eea
Combining the results, we get,
\bea
    \mathcal{F}^{\text{Haar}}_{RE} = \frac{1}{d_A^2-1} \left( 1- \frac1{d_R^2} \right) \left( f_\text{swap} -  \frac{1}{d_A}f_\text{id}\right). 
\eea

For the depolarization channel of strength $\lambda$ acting on a fraction $p$ of the qubits, we get the following expression,
\bea
\mathcal{F}^{\text{Haar}}_{RE} = \frac{d_{A}}{d_{A}^{2}-1}\left(1-\frac{1}{d_{R}^{2}}\right)\left(1-g(\lambda)^{p|A|}\right), \quad g(\lambda) = (1-\lambda)^{2}+\frac{\lambda^{2}}{3}.
\eea

If one qubit is encoded in $N$ qubits, i.e.
$d_{R} = 2$, $d_{A} = 2^{N}$, we have for $N\gg 1$,
\bea
\mathcal{F}^{\text{Haar}}_{RE} = 2^{-N+2} 3\left(1-g(\lambda)^{pN}\right).
\eea

\end{document}